\documentclass[twocolumn,aps,pra,floatfix,superscriptaddress]{revtex4-1}
\usepackage{amsfonts}
\usepackage{amssymb}
\usepackage{dcolumn}
\usepackage{graphicx}%
\usepackage{bm}
\usepackage{color}
\usepackage{braket}
\usepackage{xtab,afterpage,longtable}
\usepackage{placeins}

\makeatletter
\def\LT@LR@e{\LTleft\z@   \LTright\z@}%
\makeatother

\newcommand{\ambit}{\textsc{amb}{\footnotesize i}\textsc{t}}

\definecolor{darkgreen}{rgb}{0.0, 0.6, 0.39}
\definecolor{darkgreen2}{rgb}{0.0, 0.3, 0.19}

\begin{document}

\title{\textit{Ab initio} calculations of the spectrum of lawrencium}

\newcommand{\UNSW}{School of Physics, University of New South Wales, Sydney, New South Wales 2052, Australia
}

\newcommand{\UQ}{Australian Institute for Bioengineering and Nanotechnology, University of Queensland, Brisbane, Queensland 4072, Australia}

\newcommand{\HIM}{Helmholtz-Institut Mainz, 55128 Mainz, Germany}

\newcommand{\GSI}{GSI Helmholtzzentrum f\"ur Schwerionenforschung GmbH, 64291 Darmstadt, Germany}

\newcommand{\IFK}{Johannes Gutenberg-Universit\"at Mainz, Institut f\"ur Kernchemie, 55128 Mainz, Germany}

\newcommand{\TAU}{
School of Chemistry, Tel Aviv University, 6997801 Tel Aviv, Israel}

\newcommand{\RUG}{
Van Swinderen Institute for Particle Physics and Gravity,
University of Groningen, Nijenborgh 4, 9747 Groningen, The Netherlands\\
}

\author{E. V. Kahl}
\affiliation{\UNSW}
\affiliation{\UQ}

\author{S. Raeder}
\affiliation{\GSI}
\affiliation{\HIM}

\author{E. Eliav}
\affiliation{\TAU}

\author{A. Borschevsky}
\affiliation{\RUG}

\author{J. C. Berengut}
\affiliation{\UNSW}

\date{\today}

\begin{abstract}
%Planned optical spectroscopy experiments in lawrencium (Lr, $Z = 103$) require accurate theoretical predictions of the location of spectral lines in order to narrow the experimental search window. We present \emph{ab initio} calculations of the atomic energy levels, transition amplitudes and g-factors for lawrencium, as well as its lighter homologue lutetium (Lu, $Z = 71$). We use the configuration interaction with many-body perturbation-theory (CI+MBPT) method to calculate energy levels and transition properties, and benchmark the accuracy of our predicted energies using relativistic coupled-cluster codes. Our results are numerically converged and in close agreement with experimentally measured energy levels for Lu, and we expect a similar accuracy for Lr. These systematic calculations have identified multiple transitions of experimental utility and will serve to guide future experimental studies of Lr.

We present high accuracy relativistic investigations of the spectrum of Lr, element 103, prompted by the planned optical spectroscopy experiments on this rare and short-lived atom. Reliable predictions of the transition lines are important for the planning and success of these challenging measurements. The relativistic coupled cluster approach (CCSD(T)) was used to calculate the energies of lowest excited states, while the combination of configuration interaction method with the many-body perturbation theory (CI+MBPT) was employed to address the higher-lying states and to obtain the transition strengths and the lifetimes of the levels of experimental interest. We performed similar calculations for Lu, the lighter homologue of Lr, where experimental data is available. For the lighter element, both the calculated energies and the Einstein coefficients are in excellent agreement with the previously measured values, confirming the accuracy of the performed calculations and the reliability of our predictions for Lr. 
\end{abstract}

\maketitle

\section{Introduction}

%The optical spectroscopy of transfermium elements ($Z>100$) provides a unique test-bed for atomic and nuclear physics. 
%Experimental and theoretical studies of the electronic structure of these elements can serve to probe relativistic effects, QED corrections, and configuration-mixing effects, as well as nuclear properties via hyperfine structure measurements. 
%Recent measurements of the spectrum \cite{laatiaoui16a, raeder18a} and ionisation potential \cite{chhetri18a} of nobelium ($Z = 102$) demonstrate that precision optical spectroscopy of transfermium elements is possible even with production rates as low as a few atoms per second. 
%[id=SR, comment={one could potentially add the nobelium predictions as well}]
%The selection and narrowing of a promising search region was a key for the success of these experiments, which heavily relied on theoretical predictions.
%In this way the combination of state-of-the-art atomic structure calculations with new experimental methods has opened up new regions of the periodic table to precision optical spectroscopy.
%Lawrencium ($Z = 103$) provides a natural extension of investigations into the electronic structure of transfermium elements, with work already underway on optical spectroscopy of Lr and its ions \cite{kahl19b}. 

Optical spectroscopy of the heaviest elements can provide us with a wealth of information across various research disciplines. Such studies probe the atomic configuration and electronic structure of these atoms and give an insight into the trends in these properties, which are strongly affected by the relativistic effects \cite{Block2021, SewBacDre03, laatiaoui16a}. Predictions of chemical behaviour and material properties can also be derived from such spectroscopic studies; this is particularly important for the transfermium elements ($Z>100$), where traditional chemical studies are presently beyond our reach \cite{duellmann2017}. Information about the nuclear spin, moments, and radii can also be extracted from the measured optical spectra, complementing the nuclear decay experiments \cite{raeder18a}. Spectroscopic data for heavy elements can also be used for benchmarking the various theoretical approaches and assessing their predictive power. 

While they are both important and interesting, optical spectroscopy experiments on heaviest elements are also extremely challenging, due to the low production rates and the short lifetimes of their study subjects. Thus, alongside the specially developed ultrafast and very sensitive measurement techniques, strong theoretical support is important for the success of these experiments, particularly for narrowing the search window for the possible transitions. An example of a recent success story is the measurements of atomic levels, the hyperfine structure, and the ionization potential of nobelium \cite{laatiaoui16a,raeder18a,chhetri18a}. Theoretical predictions were important both for the success and for the interpretation of these experiments.
Further investigations of the atomic in the heaviest elements is envisaged for the low energy branch at the upcoming S$^3$ facility at GANIL, CAEN \cite{Ferrer2013,Ferrer2017towards}. 
Here laser ionization spectroscopy will be performed in a supersonic effusing gas jet, 
%a gas cell for stopping evaporation recoils, 
aiming for laser spectroscopic investigations of the heaviest elements. In addition to a search for atomic levels, Sato \textit{et al.} performed experimental measurements of the ionisation potential of Lr and the lighter actinide elements \cite{sato15a, sato18a}, which were evaluated with the support of corresponding coupled-cluster and Multi-configurational Dirac-Fock (MCDF) calculations.

A continuation of the experimental level search for heavier elements is planned at the GSI employing the RADRIS method, which was used for the successful level search in nobelium \cite{Lautenschlaeger2016115,laatiaoui16a}.
First investigations for a laser spectroscopy with this method in lawrencium have been performed, studying different filament materials \cite{Murboeck2020} to minimize the impact from surface ions in light of the fact that lawrencium has a significantly lower first ionization potential.
The application for laser spectroscopy in lawrencium is in any case challenged by a tenfold reduced production cross section compared to nobelium \cite{Gaeggeler1989}. 
In this context the reliability of theoretical predictions is important to render any successful level search possible within a realistic beam time period.

Although there are previous theoretical studies of the electronic spectrum of Lr, many of them report incomplete spectra or limited properties. 
Borschevsky \textit{et al.} \cite{borschevsky07a} used the relativistic Fock space coupled cluster approach to calculate ionisation potentials and energy levels of Lr, but that study only reported energies for levels which can be reached by exciting a single electron from the ground-state (7s$^2$7p) configuration. 
Dzuba \textit{et al.} \cite{dzuba14a} report the transition energies and the $g$-factors of Lr obtained using the configuration interaction method combined with the all-order single-double coupled-cluster technique  (CI+all order). 
MCDF was also employed for calculations of both singly- and multiply-excited states \cite{fritzsche07a}. 
However, the estimated uncertainties in those predicted excitation energies are between 1200\,cm$^{-1}$ and 2400\,cm$^{-1}$, which is too large to serve as a guide for precision spectroscopic measurements. 
Zou and Froese Fischer also carried out MCDF calculations of the energy levels and transition rates for Lu and Lr, but only calculated the three lowest lying states \cite{zou02a}.

The aim of this work is to provide accurate and reliable predictions of the level energies and the strengths of the transitions between these levels in support for the planned spectroscopy of Lr.  
%In this paper, we provide predictions of the energies and Land\`{e} g-factors for low-lying states of neutral lawrencium, as well as transition strengths and branching ratios for transitions between those states.  
The relativistic coupled-cluster method with single, double, and perturbative triple excitations (RCCSD(T)) corrected for the Breit contributions was used to calculate the lowest excitation energies following the scheme presented in Ref. \cite{PasEliBor17}. 
We also used the configuration interaction approach augmented with many-body perturbation theory (CI+MBPT) to calculate the energies of the higher-lying levels, alongside the QED contributions, g-factors and the transition rates, and to treat states that can not be handled by the coupled cluster approach. In these calculations we followed the computational scheme presented in Ref. \cite{kahl19b}. 
To facilitate the use of our predictions in the experimental context we performed extensive computational investigations that allowed us to set realistic uncertainties on our predictions. Furthermore, we have benchmarked the accuracy of our calculations by performing analogous calculations for the lighter homologue Lu, for which experimental values are available.

\section{Methods and Computational Details}

The two methods we used for the calculations of Lu and Lr energy levels are
complementary. RCCSD(T) has demonstrated high accuracy in treating heavy atoms and ions, but we can not apply this method to treat the states where an electron is excited from the 7s$^2$ shell, or states high above the ground state. These calculations are therefore restricted to the lowest configurations generated by one-electron excitations of the valence electron.  Thus, this method provides accurate energies for ``single open shell electron'' states (e.g. 7s$^2$8s, 7s$^2$6d), but completely neglects the other states (e.g. 7s7p6d). 

Consequently, we employ CI+MBPT (which has no such limitations on the number of valence electrons) to calculate the energy levels of these ``three open shell electron'' states, and as independent predictions of the ``single open shell electron'' states. 
CI+MBPT calculates atomic wavefunctions alongside
excitation energies, and we thus use this method to calculate the Land\`{e} g-factors and the Einstein
$A$-coefficients (which in turn can be used to obtain transition rates and level lifetimes).

%The calculations were carried out within the framework of the projected Dirac-Coulomb-Breit Hamiltonian \cite{Suc80} (atomic units $\hbar = m_e = e = 1$ are used throughout this work),
%\begin{eqnarray}
%H_{DCB}= \displaystyle\sum\limits_{i}h_{D}(i)+\displaystyle\sum\limits_{i<j}(1/r_{ij}+B_{ij}).
%\label{eqHdcb}
%\end{eqnarray}
%Here, $h_D$ is the one electron Dirac Hamiltonian,
%\begin{eqnarray}
%h_{D}(i)=c\, \boldsymbol \alpha_{i}\cdot \mathbf{p}_{i}+c^{2}(\beta _{i}-1)+V_\textrm{nuc}(i),
%\label{eqHd}
%\end{eqnarray}
%where $\bm{\alpha}$ and $\beta$ are the four-dimensional Dirac matrices.  The nuclear potential $V_\textrm{nuc}$  takes into account the finite size of the nucleus.  The two-electron term includes the nonrelativistic electron repulsion and the frequency independent Breit operator,
%\begin{eqnarray}
%B_{ij}=-\frac{1}{2r_{ij}}[\boldsymbol \alpha_{i}\cdot \boldsymbol \alpha_{j}+(\boldsymbol \alpha_{i}\cdot \mathbf{r}_{ij})(\boldsymbol  \alpha_{j}\cdot \mathbf{
%r}_{ij})/r_{ij}^{2}],
%\label{eqBij}
%\end{eqnarray}
%and is correct  to second order in the fine structure constant $\alpha$.

The calculations were performed within the framework of the projected Dirac-Coulomb-Breit Hamiltonian \cite{Suc80},
\begin{eqnarray}
H_{DCB}= \displaystyle\sum\limits_{i}h_{D}(i)+\displaystyle\sum\limits_{i<j}(1/r_{ij}+B_{ij}).
\label{eqHdcb}
\end{eqnarray}
Here, $h_D$ is the one electron Dirac Hamiltonian,
\begin{eqnarray}
h_{D}(i)=c\, \boldsymbol \alpha_{i}\cdot \mathbf{p}_{i}+c^{2}(\beta _{i}-1)+V_\textrm{nuc}(i),
\label{eqHd}
\end{eqnarray}
where $\bm{\alpha}$ and $\beta$ are the four-dimensional Dirac matrices.  The nuclear potential $V_\textrm{nuc}$  accounts for the finite size of the nucleus. In the RCCSD(T) calculations, the Gaussian charge distribution was used, while in the CI+MBPT calculations we used the Fermi two-parameter charge distribution model. The choice of the finite nucleus model was shown to have negligible effect on the calculated electronic properties \cite{VisDya97}.
The two-electron term includes the coulomb term and the frequency independent Breit operator,
\begin{eqnarray}
B_{ij}=-\frac{1}{2r_{ij}}[\boldsymbol \alpha_{i}\cdot \boldsymbol \alpha_{j}+(\boldsymbol \alpha_{i}\cdot \mathbf{r}_{ij})(\boldsymbol  \alpha_{j}\cdot \mathbf{
r}_{ij})/r_{ij}^{2}],
\label{eqBij}
\end{eqnarray}
and is correct to second order in the fine structure constant $\alpha$. In the RCCSD(T) calculations the Breit contribution was estimated separately and added on top of the obtained results as correction, while the CI+MBPT calculations were carried out using the explicit DCB Hamiltonian. 

\subsection{RCCSD(T)}

The RCCSD(T) calculations were carried out using the DIRAC15 program package \cite{DIRAC15}. We employed the fully uncontracted correlation-consistent all-electron relativistic basis sets of Dyall \cite{Dya06}. High quality description of the region removed from the nucleus is important for excitation energies and we have thus augmented the basis sets with a single diffuse function for each symmetry block. Finally, we extrapolated the excitation energies to the complete basis set (CBS) limit using the CBS(34) scheme \cite{HelKloKoc97}. To achieve optimal accuracy, all the electrons were correlated, and the virtual orbitals with energies below 500 \text{a.u.} were included.

The contribution of zero-frequency Breit interaction was calculated within the Fock-space coupled cluster approach (DCB-FSCC), using the Tel Aviv atomic computational package \cite{TRAFS-3C}, and added on top of the RCCSD(T) results. 
 
\subsection{CI+MBPT}

The CI+MBPT method was first developed to treat few-valence-electron atoms and ions~\cite{dzuba96a}, and as such is well-suited to the present calculations. In this method the valence electrons are treated using CI in the potential of the frozen core, while correlations with the core are treated using MBPT corrections to the radial integrals in the CI procedure. We carried out the calculations using the \ambit\ atomic structure software~\cite{kahl19a}; see also Refs.~\cite{berengut06a,berengut16a,geddes18a} for details of this implementation of the CI+MBPT method. Below we present details that are specific to the current calculations.

We start with a Dirac-Hartree-Fock (DHF) calculation in the $V^{N-1}$ potential,
including the closed-shell core plus the $n$s$^2$ electrons (where $n = 6$ for Lu and $n = 7$ for Lr). That is, all atomic electrons but one are included in the self-consistency calculations. Our DHF operator includes Breit corrections and Lamb shift corrections via the radiative potential method~\cite{flambaum05a}, which includes the self-energy \cite{ginges16a} and vacuum polarisation \cite{ginges16b} contributions. 

A large basis of single-particle orbitals, including spectroscopic and virtual orbitals, is generated by diagonalizing a set of B-splines over the DHF operator \cite{johnson88a,beloy08a,shabaev04a}. These basis orbitals are used to construct a set of many-electron configuration state functions (with well-defined angular momentum and projection) for the CI expansion. We include configurations formed by allowing all single and double excitations from the ground states (6s$^2$ 5d and 7s$^2$ 7p for Lu and Lr, respectively), up to 22spdfg (i.e. excitations to orbitals with $n < 22$, and $0 < l < 4$).

In order to reduce the size of the CI matrix and the computational load we employ the emu CI method \cite{geddes18a, kahl19a}, which exploits the fact that the wavefunctions of interest are typically dominated by contributions from a subset $N_{\mathrm{small}}$ of low-lying configurations. Off-diagonal matrix elements that do not involve at least one of these important configurations have a small effect on the eigenstates of interest, and so are set to zero without significant loss of accuracy \cite{geddes18a, dzuba17a}.
For both Lr and Lu we restrict the set of dominant configurations to those generated by taking single excitations up to 22spdfg and single and double excitations up to 12spdfg; further increasing $N_{\mathrm{small}}$
changes the energy levels by an average of 6cm$^{-1}$. In both systems, increasing the basis size beyond 22spdfg changes the energy by $\sim 1$cm$^{-1}$, indicating that the valence CI is well converged.

Core-valence correlations are included up to second order in the residual Coulomb interaction via the diagrammatic MBPT technique described in Refs.~\cite{dzuba96a,berengut06a}. We have included all 
one- and two-body diagrams with orbitals up to 35spdfgh ($n \leq 35$, $0 \leq l \leq 5$). Since our DHF includes two valence electrons, we must include subtraction diagrams in our MBPT procedure. See Refs. \cite{kahl19a, dzuba05a} and references therein for a discussion of the role of subtraction diagrams in CI+MBPT calculations. Small-scale CI+MBPT calculations showed that the choice of $V^{N-1}$ potential produces closer agreement to experimental Lu energies than other alternatives. The MBPT corrections rapidly converge as more partial waves are added, and adding orbitals with $l \geq 6$ to the MBPT basis changes the energy by an average of $108$ cm$^{-1}$.

The Land\`{e} g-factors and transition matrix elements are calculated to first-order in perturbation theory using the complete correlated CI+MBPT wavefunction. Transition lifetimes and branching ratios are derived from these matrix elements. In the CI+MBPT theory, core-valence correlations will modify the transition matrix elements. In principle, the resulting effective operators may be approximated by including higher-order corrections such as random-phase approximation~\cite{dzuba98jetp,porsev99pra,porsev01pra}. Nevertheless,
based on comparisons with experimentally measured transitions in Lu, we estimate a precision of 40\%. For Lu the experimental transition energy was used in the expression for Einstein coefficients, while for single-electron states of Lr we used  the RCCSD(T) results, and for the three-electron states the CI+MBPT energy values.

\section{Results}
\label{sec:results}

Table \ref{Lu-RCCSD(T)} contains the calculated lowest transition energies of Lu calculated using RCCSD(T). We separate contributions due to the Breit and QED effects (the latter obtained from CI+MBPT calculations with the radiative potential method), and those due to the perturbative triple excitations ($\Delta$(T)), and compare our results to the experimental values. Overall, the results are in very good agreement with the experiment. The cumulative effect of (T) and the Breit and QED corrections is quite modest, and for most transitions does not exceed 100\,cm$^{-1}$.

\begin{table*}[t]
\caption{Calculated energies of the lowest single-electron states in Lu (cm$^{-1}$).
}
\begin{ruledtabular}                                                             
\begin{tabular}{rcccccr}
%Molecule&Nucleus&Coordinate&DHF&MP2&CCSD&CCSD(T)\\\hline
 & RCCSD & (T)& $\Delta$Breit &$\Delta$QED& Final&Exp. \cite{nistasd54}\\
 \hline
 5d6s$^2$ $^2$D$_{3/2}$&   &  &   & & 0 & 0 \\
   $^2$D$_{5/2}$ &2005   & 19 & -39  & 11  & 1996$\pm 29$& 1994\\
   6s$^2$ 6p $^2$P$_{1/2}$ &  3699 &-97  &6 &50 & 3658$\pm 183$  & 4136 \\
     $^2$P$_{3/2}$ &  7287 & -264 &64  &56 &7143$\pm 298$ & 7476 \\
     6s$^2$ 7s $^2$S$_{1/2}$ &23598   &70  & 52& 79 & 23799$\pm 210$  &
     24126 \\
   %   $6s^2 7p$ $^2$P$_{3/2}$ &   &   &  &   &   &
   %  30488.62\\
 %    $6s^2 6d$ $^2$D$_{3/2}$ &   &   &  &   &   &
   %  31542.24\\
 
\end{tabular}
\end{ruledtabular}
%\begin{tablenotes}                                                   %                               \footnotesize
%\centering
%\item[\emph{a}]{
%$^\dagger$ Those results are calculated using experimental values of bond length \cite{AndZiu94}.
%}
%\end{tablenotes}
\label{Lu-RCCSD(T)}
\end{table*}

Table \ref{Lr-RCCSD(T)} contains the calculated energies of Lr. These results are in good agreement with the earlier FSCC values, but the (T) and the QED contributions are more significant in Lr (and do not cancel each other out) so the present values are expected to be more accurate. In particular $^2D$ states are very sensitive to these contributions (and to the basis set effects), which affects their values significantly.

\begin{table*}[t]
\caption{Calculated energies of the lowest single-electron states in Lr (cm$^{-1}$).
}
\begin{ruledtabular}                                                             
\begin{tabular}{rcccccr}
%Molecule&Nucleus&Coordinate&DHF&MP2&CCSD&CCSD(T)\\\hline
 & RCCSD & $\Delta$(T)& $\Delta$Breit &$\Delta$QED& Final&FSCC \cite{borschevsky07a}\\
 \hline
 7s$^2$ 7p $^2$P$_{1/2}$ &   &  &  & 0&  &0  \\
6d 7s$^2$   $^2$D$_{3/2}$& 1562   &  -416& -125  & -77  &944 $\pm 697$ & 1436 \\
    $^2$D$_{5/2}$ &  \textbf5382&-344  &-175 &-54 & 4809 $\pm 527$  &5106  \\
  7s$^2$ 7p $^2$P$_{3/2}$ & 8791 & -46 &-76  & 8 &8677$\pm 112$ & 8413  \\
     7s$^2$ 8s $^2$S$_{1/2}$ &  20284  & 291 & -82& 40 & 20533  $\pm 303$  & 20118  \\
  %    $7s^2 8p$ $^2$P$_{3/2}$ &   &   &  &   &   &27491
 %    \\
 %    $7s^2 7d$ $^2$D$_{3/2}$ &   &   &  &   &   &28096
 %     \\
 
\end{tabular}
\end{ruledtabular}
%\begin{tablenotes}                                                   %                               \footnotesize
%\centering
%\item[\emph{a}]{
%$^\dagger$ Those results are calculated using experimental values of bond length \cite{AndZiu94}.
%}
%\end{tablenotes}
\label{Lr-RCCSD(T)}
\end{table*}

An important aim of this work is to set uncertainties on the predicted transition energies, which we do for Lr by estimating the order of magnitude of the effects that are not included in the calculations. The three main remaining sources of error are the basis set incompleteness, the neglect of higher excitations beyond (T), and the higher-order QED contributions. We have extrapolated our results to the complete basis set limit and as the associated error we take the difference between the CBS result and the singly augmented ae4z (s-aug-ae4z) basis set values which is  50-500 cm$^{-1}$, depending on the transition. We assume that the effect of the higher excitations should not exceed the (T) contribution of 50-300\,cm$^{-1}$, and that the error due to the incomplete treatment of the QED effects is not larger than the vacuum polarization and the self energy contributions themselves. Note that the effect of finite nuclear size is large in Lr, of the same order as the QED contribution. However there is little uncertainty associated with this since the nuclear size can be estimated with enough accuracy.

Combining the above sources of error and assuming them to be independent, the total conservative uncertainty estimate on the calculated transition energies of Lr is given in Table II. Similar analysis was performed for Lu (Table I); the calculated transition energies generally agree with experiment within this uncertainty, supporting the validity of the proposed scheme. Note that while for Lu the dominating effects contributing to uncertainty are the basis set incompleteness and the neglect of the higher excitations, for the heavier Lr also the higher order relativistic effects become important.

Our CI+MBPT results for Lu (shown in Table \ref{tab:Lu_levels}) agree closely with experimental values from \cite{MarZalHag78}; the average disagreement between CI+MBPT and experimental energy levels is 141(294) cm$^{-1}$ (the number in brackets is the standard deviation of the difference between theory and experimental energies). We assume similar computational accuracy for Lr (Table \ref{tab:Lr_levels}) as we obtain for the lighter homologue Lu. In the absence of experimental data to compare against, we (conservatively) estimate the uncertainty in our CI+MBPT energy levels for Lr as the standard deviation of the differences between theory and experimental energy levels for Lu (294 cm$^{-1}$). We also compare our results against previous calculations from Refs. \cite{borschevsky07a, dzuba14a, fritzsche07a, zou02a}, which are presented in Table \ref{tab:lr_literature_levels}. Our results are generally in good agreement, except for the 7s$^2$6d $^2D_{3/2}$ level, where both the RCCSD(T) and CI+MBPT energies are smaller than results from previous calculations by $\sim 650$\,cm$^{-1}$. This difference can be explained by the inclusion of the perturbative triple excitations in the present work, along with a more precise treatment of the QED contributions. Both these effects lower the calculated energy.

For both RCCSD(T) and CI+MBPT, we find different ground states for Lu (6s$^2$5d) and Lr (7s$^2$7p) due to the relativistic stabilisation of the 7p orbital and anti-contraction of the 6d orbital, in agreement with earlier studies \cite{borschevsky07a,dzuba14a,fritzsche07a, zou02a}. Additionally, the 7s$^2$8s level in Lr, which is the main target state in currently planned experiments, is lower in energy by $\sim 4000$ cm$^{-1}$ than the analogous 6s$^2$7s level in Lu; again due to the relativistic contraction and stabilisation of the 8s orbital. 

\begin{figure}
    \centering
    \caption{Comparison between relativistic and nonrelativistic one-electron particle 
    density $|\Psi|^2$ for the 7s and 8s orbital in Lr}
    \includegraphics[width=0.5\textwidth]{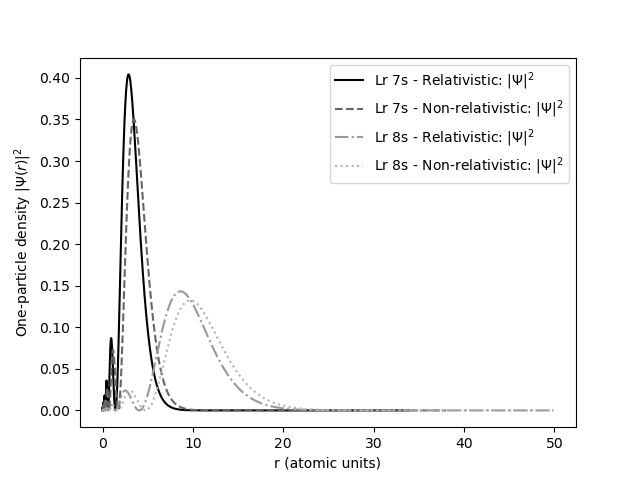}
    \label{fig:Lr_7s_orb}
\end{figure}

\begin{figure}
    \centering
    \caption{Comparison between relativistic and nonrelativistic one-electron particle 
    density $|\Psi|^2$ for the 7p orbital in Lr}
    \includegraphics[width=0.5\textwidth]{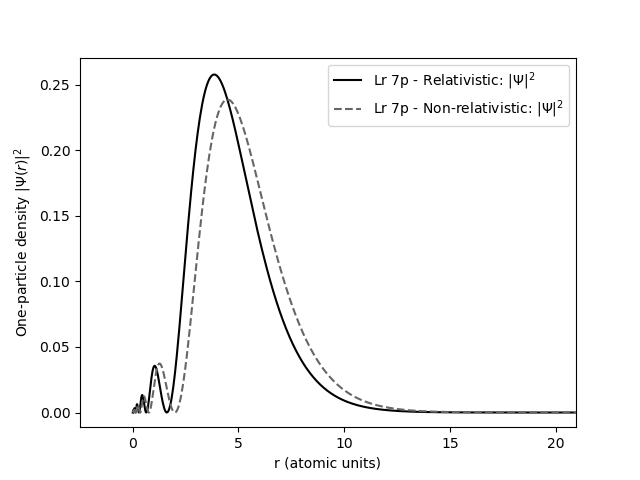}
    \label{fig:Lr_7p_orb}
\end{figure}

\begin{figure}
    \centering
    \caption{Comparison between relativistic and nonrelativistic one-electron particle 
    density $|\Psi|^2$ for the 6d orbital in Lr}
    \includegraphics[width=0.5\textwidth]{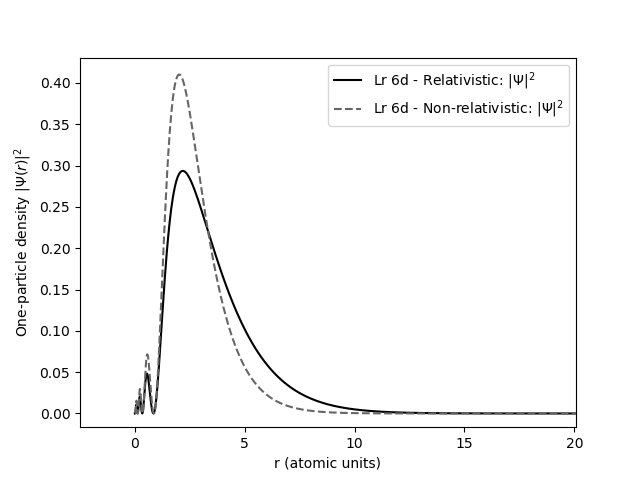}
    \label{fig:Lr_6d_orb}
\end{figure}

Figures \ref{fig:Lr_7s_orb} -- \ref{fig:Lr_6d_orb} show the effects of relativistic corrections on the radial wavefunctions of the 7s and 8s, 7p, and 6d orbitals in lawrencium. These orbitals are taken from the Dirac-Hartree-Fock step of our CI+MBPT calculations; the relativistic orbitals are the ones we have used throughout the full CI+MBPT calculations, while the nonrelativistic orbitals were obtained by setting the fine structure constant $\alpha \to 0$ (it was not possible within the structure of the \ambit\ code to do a formally nonrelativistic calculation, so the nonrelativistic orbitals presented here 
are necessarily an approximation). We can clearly see the relativistic contraction of the 7s, 8s and 7p orbitals, which in turn result in the different ordering of configurations we observe between Lu and Lr. The effect of relativity on the 6d orbital is less pronounced, as expected. 

As a result of the relativistic corrections to the orbitals in lawrencium, the 7s$^2$8s level has fewer decay channels and thus a longer lifetime of 1.46$\times 10^{-8}$s than 6s$^2$7s level in Lu, which has a lifetime of 2.42$\times 10^{-17}$s. The lifetimes of all 
Lr levels we have calculated are shown in Table \ref{tab:Lr_levels}. We included the contributions of the forbidden M1 and E2 transitions in the lifetimes of each state, but these have a negligible effect on the total lifetimes, with the exception of the 7s$^2$6d $^2D_{5/2}$ level. This level can only decay via the ``forbidden'' M1 transition to the 7s$^2$6d $^2D_{3/2}$ level, resulting a significantly longer lifetime of 0.25s compared to all other levels in Table \ref{tab:Lr_levels}.

\begin{table*}
\begin{ruledtabular}
\label{tab:Lu_levels}
\caption{Energy levels of lutetium. The CI+MBPT and RCCSD(T) columns give the energy in cm$^{-1}$ and include the 
Breit and QED corrections, the latter of which is also presented separately in the $\Delta$QED column.
Experimental values are from results tabulated in \cite{MarZalHag78}.}
\begin{tabular}{l l l l l l l l l}
Configuration	&Term	&J	&g-factor	&g-factor (expt.)	&CI+MBPT (a.u.)	&$\Delta$QED (cm$^{-1}$)
&RCCSD(T) (cm$^{-1}$)    &expt.(cm$^{-1}$)\\
\hline
6s$^2$5d 	    &$^2D$	    &3/2   &0.80	&0.79	&0	    &--	    &   &0	\\
	        &	    &5/2   &1.20	&1.20	&2174	&11   &1996	&1994	\\
\\
6s$^2$6p	    &$^2P^o$	&1/2   &0.66	&0.66	&4361	&50   &3658	&4136	\\
	        &	    &3/2   &1.33	&1.33	&7734	&56   &7143	&7182	\\
\\
5d6s6p	&$^4F^o$	&3/2	&0.45	&0.5	&17577	&-70  &	&17427	\\
	        &	    &5/2	&1.06	&1.07	&18695	&-69  &	&18505	\\
	        &	    &7/2	&1.24	&1.22	&20754	&-61  &	&20433	\\
	        &	    &9/2	&1.33	&1.3	&23016	&-55  &	&22609	\\
\\
6s5d$^2$	    &$^4F$	    &3/2	&0.41	&--	    &18324	&-98  &	&18851	\\
	        &	    &5/2	&1.03	&1.04	&18943	&-95  &	&19403	\\
	        &	    &7/2	&1.24	&--	    &19884	&-91  &	&20247	\\
	        &	    &9/2	&1.33	&--	    &21002	&-85  &	&21242	\\
\\
5d6s6p	&$^4D^o$	&1/2	&0.04	&0.00	&20783	&-72  &	&20762	\\
	        &	    &3/2	&1.16	&1.19	&21254	&-72  &	&21195	\\
	        &	    &5/2	&1.38	&1.39	&22359	&-65  &	&22222	\\
	        &	    &7/2	&1.42	&1.41	&23720	&-63  &	&23524	\\
\\
5d6s6p	&$^2D^o$	&5/2	&1.21	&1.23	&21663	&-72  &	&21462	\\
	        &	    &3/2	&0.86	&0.87	&22368	&-67  &	&22124	\\
\\
6s5d$^2$	    &$^4P$	    &1/2	&2.60	&--	    &21621	&-68  &	&21472	\\
	        &	    &3/2	&1.68	&1.73	&22618	&-73  &	&22467	\\
	        &	    &5/2	&1.43	&--	    &22842	&-81  &	&22802	\\
\\
5d6s6p	&$^4P^o$	&1/2	&2.61	&--	    &24218	&-66  &	&24108	\\
	        &	    &3/2	&1.64	&1.67	&24469	&-65  &	&24308	\\
	        &	    &5/2	&1.51	&1.53	&25501	&-60  &	&25191	\\
\\
6s$^2$7s	    &$^2S$	    &1/2	&2.02	&2.05	&24396	&79  &23799	&24126	\\
\\
6s5d$^2$	    &$^2D$	    &3/2	&0.85	&--	    &24549	&-83  &	&24518	\\
	        &	    &5/2	&1.13	&--	    &24764	&-98  &	&24711	\\
\\
6s5d$^2$	    &$^2F$	    &5/2	&1.09	&1.6	&25999	&-92  &	&25861	\\
	        &	    &7/2	&1.06	&--	    &26691	&-106  &	&26570	\\
	        &	    &9/2	&1.11	&--	    &27822	&-90  &	&26671	\\
\\
5d6s6p	&$^2F^o$	&5/2	&0.89	&0.88	&28194	&-83  &	&28020	\\
	        &	    &7/2	&1.14	&--	    &29897	&-79  &	&29487	\\
\end{tabular}
\end{ruledtabular}
\end{table*}

\begin{table*}
\caption{Energy levels of lawrencium. The CI+MBPT and RCCSD(T) columns give the energy as calculated by the two 
computational methods in cm$^{-1}$ and include the Breit and QED corrections, the latter of which (also calculated via the radiative potential method) is also
presented separately in the $\Delta$QED column. We expect that these levels will have similar accuracy to CI+MBPT results for Lu, so we (conservatively) estimate the uncertainty in our CI+MBPT energy levels for Lr as the standard deviation of the differences between theory and experimental energy levels for Lu, which is $\sim \pm 300$ cm$^{-1}$.}
\begin{ruledtabular}
\begin{tabular}{l l l l l l l l}
Configuration	&Term	&J	&g-factor	&CI+MBPT (cm$^{-1}$)	&$\Delta$QED (cm$^{-}$1)	&RCCSD(T)(cm$^{-1}$)    &Lifetime (s)\\
\hline
7s$^2$7p  &$^2P^o$	&1/2	    &0.67	    &0	    &--	    &0       &--	    \\
	    &	        &3/2	    &1.33	    &8606	&8	    &    8677   &1.79 $\times 10^{-6}$\\
\\
7s$^2$6d	&$^2D$	&3/2	    &0.80	    &712	&-77	&944        &3.69 $\times 10^{-4}$\\
	    &	        &5/2	    &1.20	    &5252	&-54	&4809       &0.25\\
\\
7s$^2$8s	&$^2S$	    &1/2	&2.00	    &20485	&40	    &   20533    &1.456$\times 10^{-8}$\\
\\
7s7p6d  &$^4F^o$	&3/2	&0.48	    &20985	&-189	&	    &1.10 $\times 10^{-7}$\\
	        &	    &5/2	&1.07	    &23289	&-188	&	    &5.89 $\times 10^{-8}$\\
	        &	    &7/2	&1.25	    &28574	&-174	&	    &2.07 $\times 10^{-7}$\\
	        &	    &9/2	&1.33	    &34758	&-169	&       &\\
\\
7s7p6d    &Odd (ambiguous)    &1/2	&0.44	    &25887	&-2	    &       &3.94 $\times 10^{-8}$\\
\\
7s7p6d	&$^4D^o$	&3/2	&1.29	    &26808	&-76	&	    &4.89 $\times 10^{-8}$\\
	        &	    &5/2	    &1.37	    &28708	&-181	&	    &6.86 $\times 10^{-8}$\\
	        &	    &7/2	    &1.35	    &33549	&-161	&	    &2.87 $\times 10^{-7}$\\
\\
7s7p$^2$	&$^4P$	        &1/2	&2.45	    &25381	&-131	&	    &3.47 $\times 10^{-8}$\\
\\
7s6d$^2$	&$^4F$	&3/2	    &0.43	    &24742	&-151	&	    &9.27 $\times 10^{-7}$\\
	    &	    &5/2	        &1.04	    &26165	&-49	&	    &1.02 $\times 10^{-5}$\\
	    &	    &7/2	        &1.23	    &28290	&-222	&	    &5.2 $\times 10^{-4}$\\
	    &	    &9/2	        &1.31	    &30754	&-212	&	    &1.36 $\times 10^{-3}$\\
\\
7s$^2$7d	&$^2D$   &3/2	        &0.80	    &28580	&31	    &	    &1.34 $\times 10^{-8}$\\
	    &   &5/2	        &1.20	    &28725	&-88	&	    &1.81 $\times 10^{-8}$\\
\\
7s$^2$8p  &$^2P^o$	&1/2	&0.39	    &26996	&-161	&	    &1.26 $\times 10^{-7}$\\
	    &	&3/2	        &1.30	    &28307	&-82	&	    &6.19 $\times 10^{-8}$\\
\\
7s$^2$9s	&$^2S$	&1/2	&2.00	    &30621	&47	    &	    &4.47 $\times 10^{-8}$\\
\\
7s$^2$9p	&$^2P^o$	&1/2  &0.86	    &32307	&-132	&	    &7.46 $\times 10^{-8}$\\
	    &	&3/2	          &1.33	    &33473	&-167	&	    &1.12 $\times 10^{-7}$\\
\\
7s$^2$6f	&$^2F^o$	&5/2  &0.86	    &31755	&-167	&	    &5.78 $\times 10^{-8}$\\
	    &	&7/2	          &1.15	    &32560	&21	    &	    &5.69 $\times 10^{-8}$\\
\end{tabular}
\end{ruledtabular}
\label{tab:Lr_levels}
\end{table*}

\begin{table*}
\caption{Comparison of Lr energy levels with previous calculations. All energies are given in cm$^{-1}$; 
energy levels not present in a particular source are left blank.}
\begin{ruledtabular}
\begin{tabular}{l l l l l l l l l}
Configuration	&Term	&J	&g-factor	&CI+MBPT	&FSCC\cite{borschevsky07a}	
&CI + all order \cite{dzuba14a} &MCDF \cite{fritzsche07a}	&MCDF\cite{zou02a}	\\
\hline
7s$^2$7p	&$^2P^o$    &1/2	 &2.01	    &0	    &0	    &0	    &0	    &0	\\
	&	            &3/2	 &0.80	&8606	&8413	&8495	&8138	&7807\\
\\
7s$^2$6d	&$^2D$	    &3/2 &0.80	&712	&1436	&1555	&1331	&1127\\
	&	            &5/2	 &1.20	&5252	&5106	&5423	&4187	&	\\
\\
7s$^2$8s	&$^2S$	        &1/2	&2.01	&20485	&20118	&20253	&20405	&	\\
\\
7s7p6d	&$^4F^o$       &3/2	 &0.43	&20985	&	    &21288	&20886	&	\\
	&	            &5/2	 &1.20	&23289	&       &23530	&23155	&	\\
	&	            &7/2	 &1.23	&28574	&	    &28320	&27276	&	\\
	&	            &9/2	 &1.31	&34758	&	    &34212	&32775	&	\\
\\
7s7p6d	&Odd (ambiguous)	&1/2	&2.44	&25887	&	    &	    &27904	&	\\
\\
7s7p6d	&$^4D^o$	&3/2	 &0.83	&26808	&	&	&	&	\\
	&	            &5/2	 &1.04	&28708	&	&	&	&	\\
	&	            &7/2	 &0.88	&33549	&	&	&	&	\\
\\
7s7p$^2$	&$^4P$      &1/2	 &2.44	&25380	&	&	&	&	\\
\\
7s6d$^2$	&$^4F$	&3/2	    &0.43	&24742	&	&	&25409	&	\\
	&	            &5/2	    &1.04	&26165	&	&	&27397	&	\\
	&	            &7/2	    &1.23	&28290	&	&	&	&	\\
	&	            &9/2	    &1.31	&30754	&	&	&34807	&	\\
\\
7s$^2$7d	&$^2D$	&3/2	    &0.83	&28580	&28118	&	&	&	\\
	&	            &5/2	    &1.22	&28725	&28385	&	&	&	\\
\\
7s$^2$8p	&$^2P^o$&1/2	 &2.00	&26996	&26111	&25912	&25246	&	\\
	&	            &3/2	 &1.61	&28307	&27508	&27079	&26902	&	\\
\\
7s$^2$9s	&$^2S$	&1/2	 &2.00	&30621	&30119	&	&	&	\\
\\
7s$^2$9p	&$^2P^o$&1/2	 &2.21	&32307	&32295	&	&	&	\\
	&	            &3/2	 &0.80	&33473	&32840	&	&	&	\\
\\
7s$^2$6f	&$^2F^o$&5/2	 &1.37	&31755	&32949	&	&	&	\\
	&	            &7/2	 &0.92	&32560	&32950	&	&	&	\\
\end{tabular}
\label{tab:lr_literature_levels}
\end{ruledtabular}
\end{table*}

\begin{figure*}
    \centering
    \includegraphics[width=\textwidth]{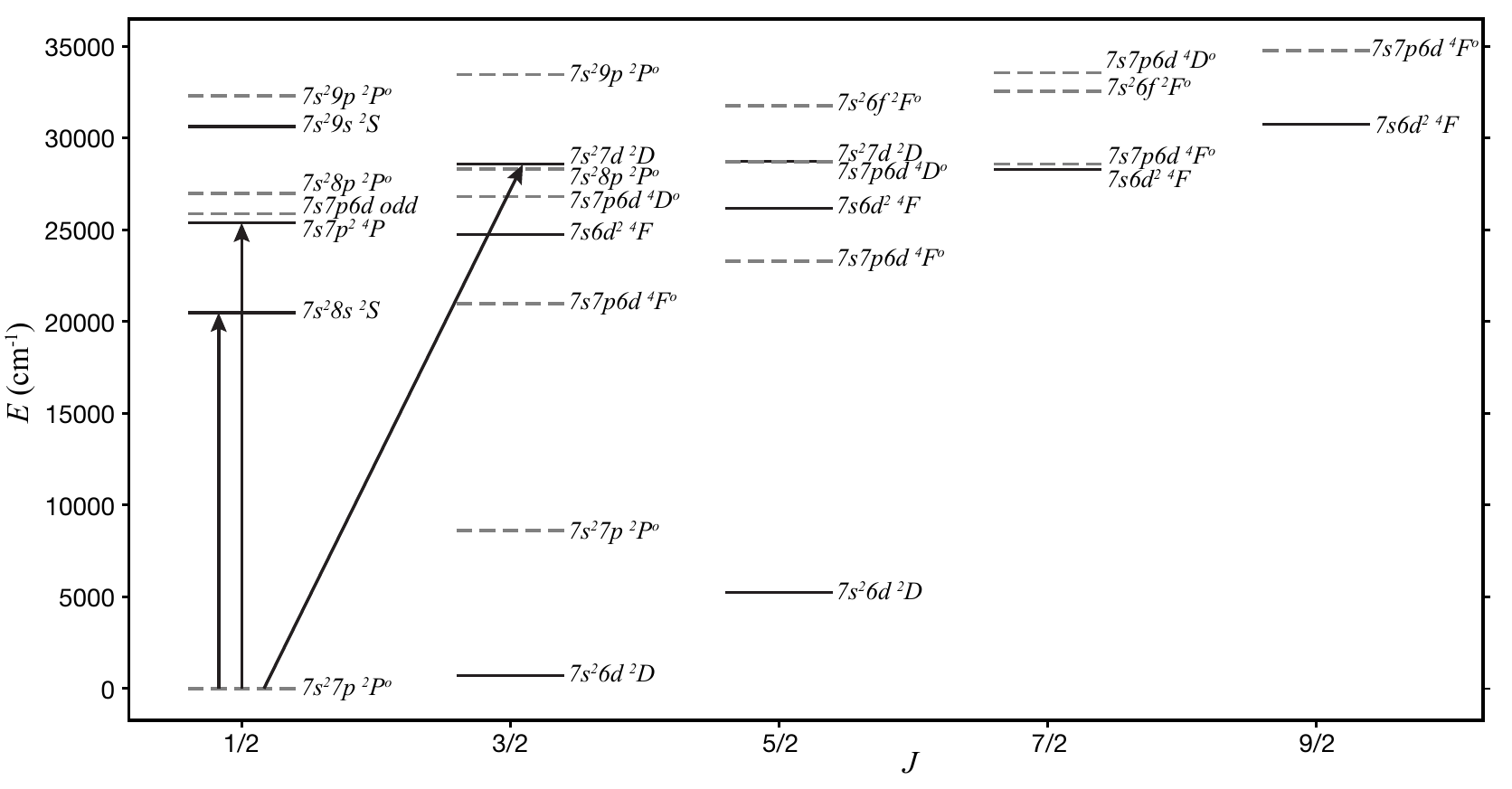}
    \caption{Grotrian diagram of the energy level scheme of Lr, as calculated by the CI+MBPT method including Breit and QED corrections. Arrows represent transitions that may be targeted in experiment, as discussed in the text.}
    \label{fig:lr_grotrian}
\end{figure*}

Table \ref{tab:Lu_transitions} shows the Einstein $A$-coefficients (transition probabilities) for the
lowest-lying electric dipole (E1) transitions in lutetium. The spectra of both Lu and Lr are 
relatively dense, so to reduce the size of the tables we have only included Lu transitions which 
fulfill two criteria: they must have experimentally derived $A$-coefficients tabulated in ref. 
\cite{nistasd54}, and at least one state in the transition must be analogous to the target states 
in Lr. Our $A$-coefficients are mostly larger than the experimentally derived values and differ from 
experimental results by an average of $40\%$, although transitions involving states with $J \geq 2.5$ tend 
to have worse accuracy. The relative strengths of the different transitions are reproduced and we can 
reliably identify the strongest transitions.

We present electric-dipole transition energies and $A$-coefficients between low-lying and target levels of 
lawrencium in table  \ref{tab:E1_transitions}, where we expect similar accuracy to the E1 transitions in 
Lu. As in table \ref{tab:Lu_transitions}, we have only included results for the levels which are of 
experimental interest, as well as states lower in energy than the experimental targets. The forbidden M1 
and E2 transitions contribute a negligible amount compared to the dominant E1 transitions (except for the 
aforementioned 7s$^2$6d $^2D_{5/2}$ level, which can only decay via M1 transition), so we have not included
them in the tables.

\section{Effect of finite nuclear size}

All of our calculations use a finite-size nuclear model with root-mean-square radius $R_{rms} = 6.00$~fm. It may be possible in to extract difference in nuclear charge radii between different isotopes of Lr by studying the isotope shift. In heavy atoms the isotope shift of a transition $i$ is dominated by the field shift, which is usually parameterized by
\begin{equation}
\label{eq:F}
\delta \nu_i^{A,A'} = F_i\, \delta \langle R_{rms}^2\rangle^{A,A'}
\end{equation}
where $\delta \langle R_{rms}^2\rangle^{A,A'}$ is the change in mean-square radius between the isotopes $A$ and $A'$.

\begin{figure}
\centering
\includegraphics[width=0.5\textwidth]{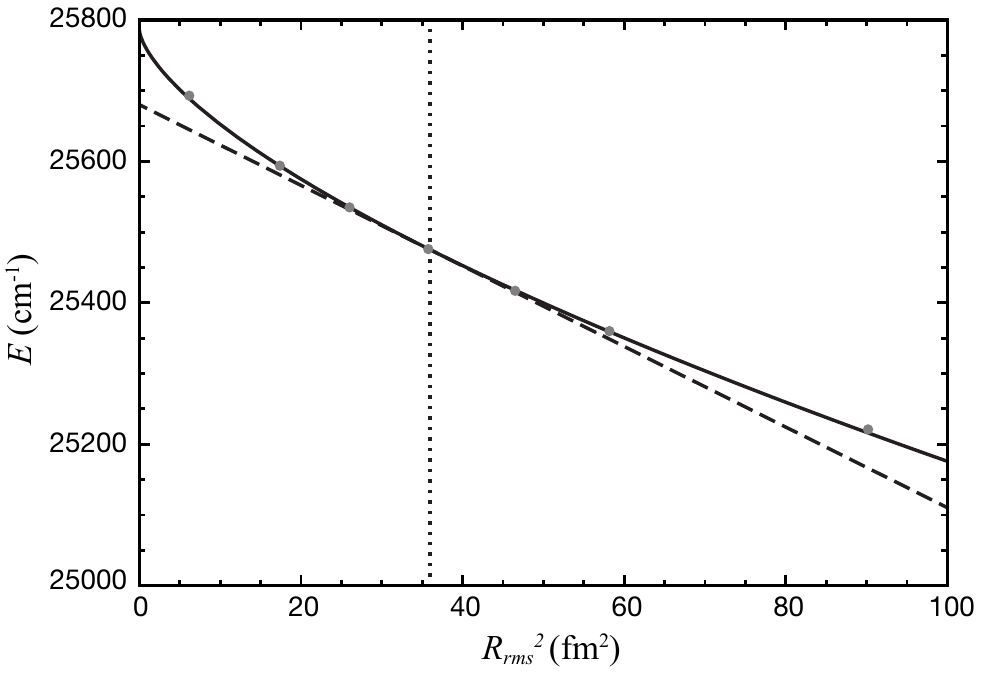}
\caption{Effect of finite nuclear size on the $7s^2\;7p\ ^2P_{1/2}^o \rightarrow 7s\;7p^2\ ^4P_{1/2}$ transition energy. Grey points: CI+MBPT calculation; solid line: linear fit in $R_{rms}^{2\gamma}$; dashed line: linear fit in $R_{rms}^2$ showing usual field shift parameterization which is tangent to solid line at $R_{rms} = 6$~fm (dotted line).}
\label{fig:FvsR2}
\end{figure}

For highly relativistic systems such as Lr, the relationship is better expressed by (see, e.g.~\cite{sobelman72book,dzuba17pra0})
\begin{equation}
\label{eq:Ftilde}
\delta \nu_i^{A,A'} = \tilde F_i\, \delta \langle R_{rms}^{2\gamma} \rangle^{A,A'}
\end{equation}
where $\gamma = \sqrt{1 - (Z\alpha)^2} \approx 0.66$ for Lr. In Figure~\ref{fig:FvsR2} we show CI+MBPT calculations for the $7s^2\;7p\ ^2P_{1/2}^o \rightarrow 7s\;7p^2\ ^4P_{1/2}$ transition in Lr, which is quite sensitive to finite nuclear size due to the change in $s$-wave electron number. We also show fits using (\ref{eq:F}) and (\ref{eq:Ftilde}). These clearly show that the latter parameterization remains more accurate over a larger range of nuclear radius. We see that even in this case, the total effect of finite nuclear size is around 300~cm$^{-1}$, which is the same order as the uncertainty in our calculations. The field shift constants $F$ and $\tilde F$ are, however, quite stable and could be used to extract changes in $R_{rms}$ between isotopes at the $\sim 10$\% level or better. We have calculated these values for the transitions of experimental interest and presented them in Table~\ref{tab:F_values}.

\begin{table}[htb]
\begin{ruledtabular}
\begin{tabular}{lcc}
Upper level & $F$ (cm$^{-1}$/fm$^2$) & $\tilde F$ (cm$^{-1}$/fm$^{2\gamma}$) \\
\hline
$7s^2\;8s\ ^2S_{1/2}$ &  1.507 &   7.740 \\
$7s\;7p^2\ ^4P_{1/2}$ & -5.699 & -29.263 \\
$7s^2\;7d\ ^2D_{3/2}$ &  0.953 &   4.894 \\
\end{tabular}
\caption{Field shift constants for ground state transitions of experimental interest in Lr. $F$ and $\tilde F$ are defined by Eqs. (\ref{eq:F}) and (\ref{eq:Ftilde}), respectively.}
\label{tab:F_values}
\end{ruledtabular}
\end{table}

\section{Summary and conclusion}

From the presented results it can be concluded that both RCCSD(T) and CI+MBPT methods are able to reliably calculate the experimental spectrum of neutral Lu with good accuracy. 
Moreover these methods give energies of Lr that are in good agreement with each other. 
Therefore we have confidence in the accuracy of both approaches for calculating low-lying states of Lr which will enable a proper planning of any experimental search for atomic levels in Lr, while additional properties of the individual levels such as the transition strengths and lifetimes will help in the validation of the configuration once a resonance is identified. 

The levels calculated in this work indicate three transitions from the atomic ground state with a suitable transition strengths with Einstein $A$-coefficients above $10^7$\,s$^{-1}$ which is required to ensure an efficient transfer of the population (see Fig.~\ref{fig:lr_grotrian}). 
The transitions target the excited  7s$^2$\,8s~$^2S_{1/2}$ level at 20\,485\,cm$^{-1}$ with a transition strength of $3.31\times 10^{7}$\,s$^{-1}$, the excited  7s\,7p$^2$~$^4P_{1/2}$ level at 25\,381\,cm$^{-1}$ with a transition strength of $2.51\times 10^{7}$\,s$^{-1}$ and the excited  7s$^2$\,7d$~^2D_{3/2}$ level at 28\,580\,cm$^{-1}$ with a transition strength of $6.14\times 10^{7}$\,s$^{-1}$. 
For laser spectroscopy regarding the extraction of nuclear properties the latter transition to the 28\,580\,cm$^{-1}$ state with $J=3/2$ is beneficial due to a sensitivity to the nuclear spectroscopic quadrupole moment \cite{Block2021}.

However, from the calculations it becomes evident that three very close levels with lower energy ( 7s7d6d~$^4F^{\circ}_{7/2}$ at 28\,574\,cm$^{-1}$,
7s$^2$8p~$^2P^{\circ}_{3/2}$ at 28\,307\,cm$^{-1}$ and 7s6d$^2$~$^4F^{\circ}_{7/2}$ at 28\,290\,cm$^{-1}$ ) are present in the atomic structure of Lr\,I.
A fast quenching into these states induced by the buffer gas used in the RADRIS experiment is quite possible and was already observed for a similar energy difference of atomic levels in nobelium \cite{laatiaoui16a, chhetri17}.
As the lifetimes of these states are similar it will be difficult to distinguish the levels by successive, delayed excitation or ionization. 
Although this might disturb a measurement of the ionization potential when addressing higher lying Rydberg levels from this state, the measurement of a hyperfine structure is not affected as the depopulation from the quenching only occurs after the optical excitation.
For the determination of the first ionization potential by Rydberg convergence, as done in the case of No \cite{chhetri18a}, the excited  7s$^2$\,8s$~^2S_{1/2}$ level at 20\,485\,cm$^{-1}$ and the excited  7s\,7p$^2~^4P_{1/2}$ level at 25\,381\,cm$^{-1}$ are promising, as they are well separated from other atomic levels, preventing buffer gas induced quenching effects and by the fact that the expected energy difference to the ionization potential can be addressed with the available laser systems.

\acknowledgements

The work of EVK was supported by the Australian Government Research Training Program scholarship. 
The authors would like to thank the Center for Information Technology of the University of Groningen for providing access to the Peregrine high performance
computing cluster and for their technical support.
We acknowledge Research Technology Services at UNSW Sydney for supporting this project with additional computing resources.
This research was undertaken with the assistance of resources and services from the National Computational Infrastructure (NCI), which is supported by the Australian Government.
This work was supported by the Australian Research Council (Grant No. DP190100974).
This project has received funding from the European Union's Horizon 2020 research and innovation programme under grant agreement No 861198 — LISA — H2020-MSCA-ITN-2019.
AB is grateful for the support of the UNSW Gordon Godfrey fellowship.
\bibliography{Lr}

\begin{thebibliography}{46}%
\makeatletter
\providecommand \@ifxundefined [1]{%
 \@ifx{#1\undefined}
}%
\providecommand \@ifnum [1]{%
 \ifnum #1\expandafter \@firstoftwo
 \else \expandafter \@secondoftwo
 \fi
}%
\providecommand \@ifx [1]{%
 \ifx #1\expandafter \@firstoftwo
 \else \expandafter \@secondoftwo
 \fi
}%
\providecommand \natexlab [1]{#1}%
\providecommand \enquote  [1]{``#1''}%
\providecommand \bibnamefont  [1]{#1}%
\providecommand \bibfnamefont [1]{#1}%
\providecommand \citenamefont [1]{#1}%
\providecommand \href@noop [0]{\@secondoftwo}%
\providecommand \href [0]{\begingroup \@sanitize@url \@href}%
\providecommand \@href[1]{\@@startlink{#1}\@@href}%
\providecommand \@@href[1]{\endgroup#1\@@endlink}%
\providecommand \@sanitize@url [0]{\catcode `\\12\catcode `\$12\catcode
  `\&12\catcode `\#12\catcode `\^12\catcode `\_12\catcode `\%12\relax}%
\providecommand \@@startlink[1]{}%
\providecommand \@@endlink[0]{}%
\providecommand \url  [0]{\begingroup\@sanitize@url \@url }%
\providecommand \@url [1]{\endgroup\@href {#1}{\urlprefix }}%
\providecommand \urlprefix  [0]{URL }%
\providecommand \Eprint [0]{\href }%
\providecommand \doibase [0]{http://dx.doi.org/}%
\providecommand \selectlanguage [0]{\@gobble}%
\providecommand \bibinfo  [0]{\@secondoftwo}%
\providecommand \bibfield  [0]{\@secondoftwo}%
\providecommand \translation [1]{[#1]}%
\providecommand \BibitemOpen [0]{}%
\providecommand \bibitemStop [0]{}%
\providecommand \bibitemNoStop [0]{.\EOS\space}%
\providecommand \EOS [0]{\spacefactor3000\relax}%
\providecommand \BibitemShut  [1]{\csname bibitem#1\endcsname}%
\let\auto@bib@innerbib\@empty
%</preamble>
\bibitem [{\citenamefont {Block}\ \emph {et~al.}(2021)\citenamefont {Block},
  \citenamefont {Laatiaoui},\ and\ \citenamefont {Raeder}}]{Block2021}%
  \BibitemOpen
  \bibfield  {author} {\bibinfo {author} {\bibfnamefont {M.}~\bibnamefont
  {Block}}, \bibinfo {author} {\bibfnamefont {M.}~\bibnamefont {Laatiaoui}}, \
  and\ \bibinfo {author} {\bibfnamefont {S.}~\bibnamefont {Raeder}},\ }\href
  {\doibase 10.1016/j.ppnp.2020.103834} {\bibfield  {journal} {\bibinfo
  {journal} {Progress in Particle and Nuclear Physics}\ }\textbf {\bibinfo
  {volume} {116}},\ \bibinfo {pages} {103834} (\bibinfo {year}
  {2021})}\BibitemShut {NoStop}%
\bibitem [{\citenamefont {Sewtz}\ \emph {et~al.}(2003)\citenamefont {Sewtz},
  \citenamefont {Backe}, \citenamefont {Dretzke}, \citenamefont {Kube},
  \citenamefont {Lauth}, \citenamefont {Schwamb}, \citenamefont {Eberhardt},
  \citenamefont {Gr\"uning}, \citenamefont {Th\"orle}, \citenamefont
  {Trautmann}, \citenamefont {Kunz}, \citenamefont {Lassen}, \citenamefont
  {Passler}, \citenamefont {Dong}, \citenamefont {Fritzsche},\ and\
  \citenamefont {Haire}}]{SewBacDre03}%
  \BibitemOpen
  \bibfield  {author} {\bibinfo {author} {\bibfnamefont {M.}~\bibnamefont
  {Sewtz}}, \bibinfo {author} {\bibfnamefont {H.}~\bibnamefont {Backe}},
  \bibinfo {author} {\bibfnamefont {A.}~\bibnamefont {Dretzke}}, \bibinfo
  {author} {\bibfnamefont {G.}~\bibnamefont {Kube}}, \bibinfo {author}
  {\bibfnamefont {W.}~\bibnamefont {Lauth}}, \bibinfo {author} {\bibfnamefont
  {P.}~\bibnamefont {Schwamb}}, \bibinfo {author} {\bibfnamefont
  {K.}~\bibnamefont {Eberhardt}}, \bibinfo {author} {\bibfnamefont
  {C.}~\bibnamefont {Gr\"uning}}, \bibinfo {author} {\bibfnamefont
  {P.}~\bibnamefont {Th\"orle}}, \bibinfo {author} {\bibfnamefont
  {N.}~\bibnamefont {Trautmann}}, \bibinfo {author} {\bibfnamefont
  {P.}~\bibnamefont {Kunz}}, \bibinfo {author} {\bibfnamefont {J.}~\bibnamefont
  {Lassen}}, \bibinfo {author} {\bibfnamefont {G.}~\bibnamefont {Passler}},
  \bibinfo {author} {\bibfnamefont {C.~Z.}\ \bibnamefont {Dong}}, \bibinfo
  {author} {\bibfnamefont {S.}~\bibnamefont {Fritzsche}}, \ and\ \bibinfo
  {author} {\bibfnamefont {R.~G.}\ \bibnamefont {Haire}},\ }\href {\doibase
  10.1103/PhysRevLett.90.163002} {\bibfield  {journal} {\bibinfo  {journal}
  {Phys. Rev. Lett.}\ }\textbf {\bibinfo {volume} {90}},\ \bibinfo {pages}
  {163002} (\bibinfo {year} {2003})}\BibitemShut {NoStop}%
\bibitem [{\citenamefont {Laatiaoui}\ \emph {et~al.}(2016)\citenamefont
  {Laatiaoui}, \citenamefont {Lauth}, \citenamefont {Backe}, \citenamefont
  {Block}, \citenamefont {Ackermann}, \citenamefont {Cheal}, \citenamefont
  {Chhetri}, \citenamefont {D\"{u}llmann}, \citenamefont {van Duppen},
  \citenamefont {Even}, \citenamefont {Ferrer}, \citenamefont {Giacoppo},
  \citenamefont {G\"{o}tz}, \citenamefont {He\"sberger}, \citenamefont {Huyse},
  \citenamefont {Kaleja}, \citenamefont {Khuyagbaatar}, \citenamefont {Kunz},
  \citenamefont {Lautenschl\"{a}ger}, \citenamefont {Mistry}, \citenamefont
  {Raeder}, \citenamefont {Ramirez}, \citenamefont {Walther}, \citenamefont
  {Wraith},\ and\ \citenamefont {Yakushev}}]{laatiaoui16a}%
  \BibitemOpen
  \bibfield  {author} {\bibinfo {author} {\bibfnamefont {M.}~\bibnamefont
  {Laatiaoui}}, \bibinfo {author} {\bibfnamefont {W.}~\bibnamefont {Lauth}},
  \bibinfo {author} {\bibfnamefont {H.}~\bibnamefont {Backe}}, \bibinfo
  {author} {\bibfnamefont {M.}~\bibnamefont {Block}}, \bibinfo {author}
  {\bibfnamefont {D.}~\bibnamefont {Ackermann}}, \bibinfo {author}
  {\bibfnamefont {B.}~\bibnamefont {Cheal}}, \bibinfo {author} {\bibfnamefont
  {P.}~\bibnamefont {Chhetri}}, \bibinfo {author} {\bibfnamefont {C.~E.}\
  \bibnamefont {D\"{u}llmann}}, \bibinfo {author} {\bibfnamefont
  {P.}~\bibnamefont {van Duppen}}, \bibinfo {author} {\bibfnamefont
  {J.}~\bibnamefont {Even}}, \bibinfo {author} {\bibfnamefont {R.}~\bibnamefont
  {Ferrer}}, \bibinfo {author} {\bibfnamefont {F.}~\bibnamefont {Giacoppo}},
  \bibinfo {author} {\bibfnamefont {S.}~\bibnamefont {G\"{o}tz}}, \bibinfo
  {author} {\bibfnamefont {F.~P.}\ \bibnamefont {He\"sberger}}, \bibinfo
  {author} {\bibfnamefont {M.}~\bibnamefont {Huyse}}, \bibinfo {author}
  {\bibfnamefont {O.}~\bibnamefont {Kaleja}}, \bibinfo {author} {\bibfnamefont
  {J.}~\bibnamefont {Khuyagbaatar}}, \bibinfo {author} {\bibfnamefont
  {P.}~\bibnamefont {Kunz}}, \bibinfo {author} {\bibfnamefont {F.}~\bibnamefont
  {Lautenschl\"{a}ger}}, \bibinfo {author} {\bibfnamefont {A.~K.}\ \bibnamefont
  {Mistry}}, \bibinfo {author} {\bibfnamefont {S.}~\bibnamefont {Raeder}},
  \bibinfo {author} {\bibfnamefont {E.~M.}\ \bibnamefont {Ramirez}}, \bibinfo
  {author} {\bibfnamefont {T.}~\bibnamefont {Walther}}, \bibinfo {author}
  {\bibfnamefont {C.}~\bibnamefont {Wraith}}, \ and\ \bibinfo {author}
  {\bibfnamefont {A.}~\bibnamefont {Yakushev}},\ }\href@noop {} {\bibfield
  {journal} {\bibinfo  {journal} {Nature}\ }\textbf {\bibinfo {volume} {538}}
  (\bibinfo {year} {2016})}\BibitemShut {NoStop}%
\bibitem [{\citenamefont {D{\"u}llmann}(2017)}]{duellmann2017}%
  \BibitemOpen
  \bibfield  {author} {\bibinfo {author} {\bibfnamefont {C.~E.}\ \bibnamefont
  {D{\"u}llmann}},\ }\href {\doibase 10.1080/10619127.2017.1280333} {\bibfield
  {journal} {\bibinfo  {journal} {Nuclear Physics News}\ }\textbf {\bibinfo
  {volume} {27}},\ \bibinfo {pages} {14} (\bibinfo {year} {2017})}\BibitemShut
  {NoStop}%
\bibitem [{\citenamefont {Raeder}\ \emph {et~al.}(2018)\citenamefont {Raeder},
  \citenamefont {Ackermann}, \citenamefont {Backe}, \citenamefont {Beerwerth},
  \citenamefont {Berengut}, \citenamefont {Block}, \citenamefont {Borschevsky},
  \citenamefont {Cheal}, \citenamefont {Chhetri}, \citenamefont {D\"{u}llmann},
  \citenamefont {Dzuba}, \citenamefont {Eliav}, \citenamefont {Even},
  \citenamefont {Ferrer}, \citenamefont {Flambaum}, \citenamefont {Fritzsche},
  \citenamefont {Giacoppo}, \citenamefont {G\"otz}, \citenamefont
  {He\ss{}berger}, \citenamefont {Huyse}, \citenamefont {Kaldor}, \citenamefont
  {Kaleja}, \citenamefont {Khuyagbaatar}, \citenamefont {Kunz}, \citenamefont
  {Laatiaoui}, \citenamefont {Lautenschl\"ager}, \citenamefont {Lauth},
  \citenamefont {Mistry}, \citenamefont {Minaya~Ramirez}, \citenamefont
  {Nazarewicz}, \citenamefont {Porsev}, \citenamefont {Safronova},
  \citenamefont {Safronova}, \citenamefont {Schuetrumpf}, \citenamefont
  {Van~Duppen}, \citenamefont {Walther}, \citenamefont {Wraith},\ and\
  \citenamefont {Yakushev}}]{raeder18a}%
  \BibitemOpen
  \bibfield  {author} {\bibinfo {author} {\bibfnamefont {S.}~\bibnamefont
  {Raeder}}, \bibinfo {author} {\bibfnamefont {D.}~\bibnamefont {Ackermann}},
  \bibinfo {author} {\bibfnamefont {H.}~\bibnamefont {Backe}}, \bibinfo
  {author} {\bibfnamefont {R.}~\bibnamefont {Beerwerth}}, \bibinfo {author}
  {\bibfnamefont {J.~C.}\ \bibnamefont {Berengut}}, \bibinfo {author}
  {\bibfnamefont {M.}~\bibnamefont {Block}}, \bibinfo {author} {\bibfnamefont
  {A.}~\bibnamefont {Borschevsky}}, \bibinfo {author} {\bibfnamefont
  {B.}~\bibnamefont {Cheal}}, \bibinfo {author} {\bibfnamefont
  {P.}~\bibnamefont {Chhetri}}, \bibinfo {author} {\bibfnamefont {C.~E.}\
  \bibnamefont {D\"{u}llmann}}, \bibinfo {author} {\bibfnamefont {V.~A.}\
  \bibnamefont {Dzuba}}, \bibinfo {author} {\bibfnamefont {E.}~\bibnamefont
  {Eliav}}, \bibinfo {author} {\bibfnamefont {J.}~\bibnamefont {Even}},
  \bibinfo {author} {\bibfnamefont {R.}~\bibnamefont {Ferrer}}, \bibinfo
  {author} {\bibfnamefont {V.~V.}\ \bibnamefont {Flambaum}}, \bibinfo {author}
  {\bibfnamefont {S.}~\bibnamefont {Fritzsche}}, \bibinfo {author}
  {\bibfnamefont {F.}~\bibnamefont {Giacoppo}}, \bibinfo {author}
  {\bibfnamefont {S.}~\bibnamefont {G\"otz}}, \bibinfo {author} {\bibfnamefont
  {F.~P.}\ \bibnamefont {He\ss{}berger}}, \bibinfo {author} {\bibfnamefont
  {M.}~\bibnamefont {Huyse}}, \bibinfo {author} {\bibfnamefont
  {U.}~\bibnamefont {Kaldor}}, \bibinfo {author} {\bibfnamefont
  {O.}~\bibnamefont {Kaleja}}, \bibinfo {author} {\bibfnamefont
  {J.}~\bibnamefont {Khuyagbaatar}}, \bibinfo {author} {\bibfnamefont
  {P.}~\bibnamefont {Kunz}}, \bibinfo {author} {\bibfnamefont {M.}~\bibnamefont
  {Laatiaoui}}, \bibinfo {author} {\bibfnamefont {F.}~\bibnamefont
  {Lautenschl\"ager}}, \bibinfo {author} {\bibfnamefont {W.}~\bibnamefont
  {Lauth}}, \bibinfo {author} {\bibfnamefont {A.~K.}\ \bibnamefont {Mistry}},
  \bibinfo {author} {\bibfnamefont {E.}~\bibnamefont {Minaya~Ramirez}},
  \bibinfo {author} {\bibfnamefont {W.}~\bibnamefont {Nazarewicz}}, \bibinfo
  {author} {\bibfnamefont {S.~G.}\ \bibnamefont {Porsev}}, \bibinfo {author}
  {\bibfnamefont {M.~S.}\ \bibnamefont {Safronova}}, \bibinfo {author}
  {\bibfnamefont {U.~I.}\ \bibnamefont {Safronova}}, \bibinfo {author}
  {\bibfnamefont {B.}~\bibnamefont {Schuetrumpf}}, \bibinfo {author}
  {\bibfnamefont {P.}~\bibnamefont {Van~Duppen}}, \bibinfo {author}
  {\bibfnamefont {T.}~\bibnamefont {Walther}}, \bibinfo {author} {\bibfnamefont
  {C.}~\bibnamefont {Wraith}}, \ and\ \bibinfo {author} {\bibfnamefont
  {A.}~\bibnamefont {Yakushev}},\ }\href {\doibase
  10.1103/PhysRevLett.120.232503} {\bibfield  {journal} {\bibinfo  {journal}
  {Phys. Rev. Lett.}\ }\textbf {\bibinfo {volume} {120}},\ \bibinfo {pages}
  {232503} (\bibinfo {year} {2018})}\BibitemShut {NoStop}%
\bibitem [{\citenamefont {Chhetri}\ \emph {et~al.}(2018)\citenamefont
  {Chhetri}, \citenamefont {Ackermann}, \citenamefont {Backe}, \citenamefont
  {Block}, \citenamefont {Cheal}, \citenamefont {Droese}, \citenamefont
  {D\"{u}llmann}, \citenamefont {Even}, \citenamefont {Ferrer}, \citenamefont
  {Giacoppo}, \citenamefont {G\"{o}tz}, \citenamefont {He\ss{}berger},
  \citenamefont {Huyse}, \citenamefont {Kaleja}, \citenamefont {Khuyagbaatar},
  \citenamefont {Kunz}, \citenamefont {Laatiaoui}, \citenamefont
  {Lautenschl\"ager}, \citenamefont {Lauth}, \citenamefont {Lecesne},
  \citenamefont {Lens}, \citenamefont {Minaya~Ramirez}, \citenamefont {Mistry},
  \citenamefont {Raeder}, \citenamefont {Van~Duppen}, \citenamefont {Walther},
  \citenamefont {Yakushev},\ and\ \citenamefont {Zhang}}]{chhetri18a}%
  \BibitemOpen
  \bibfield  {author} {\bibinfo {author} {\bibfnamefont {P.}~\bibnamefont
  {Chhetri}}, \bibinfo {author} {\bibfnamefont {D.}~\bibnamefont {Ackermann}},
  \bibinfo {author} {\bibfnamefont {H.}~\bibnamefont {Backe}}, \bibinfo
  {author} {\bibfnamefont {M.}~\bibnamefont {Block}}, \bibinfo {author}
  {\bibfnamefont {B.}~\bibnamefont {Cheal}}, \bibinfo {author} {\bibfnamefont
  {C.}~\bibnamefont {Droese}}, \bibinfo {author} {\bibfnamefont {C.~E.}\
  \bibnamefont {D\"{u}llmann}}, \bibinfo {author} {\bibfnamefont
  {J.}~\bibnamefont {Even}}, \bibinfo {author} {\bibfnamefont {R.}~\bibnamefont
  {Ferrer}}, \bibinfo {author} {\bibfnamefont {F.}~\bibnamefont {Giacoppo}},
  \bibinfo {author} {\bibfnamefont {S.}~\bibnamefont {G\"{o}tz}}, \bibinfo
  {author} {\bibfnamefont {F.~P.}\ \bibnamefont {He\ss{}berger}}, \bibinfo
  {author} {\bibfnamefont {M.}~\bibnamefont {Huyse}}, \bibinfo {author}
  {\bibfnamefont {O.}~\bibnamefont {Kaleja}}, \bibinfo {author} {\bibfnamefont
  {J.}~\bibnamefont {Khuyagbaatar}}, \bibinfo {author} {\bibfnamefont
  {P.}~\bibnamefont {Kunz}}, \bibinfo {author} {\bibfnamefont {M.}~\bibnamefont
  {Laatiaoui}}, \bibinfo {author} {\bibfnamefont {F.}~\bibnamefont
  {Lautenschl\"ager}}, \bibinfo {author} {\bibfnamefont {W.}~\bibnamefont
  {Lauth}}, \bibinfo {author} {\bibfnamefont {N.}~\bibnamefont {Lecesne}},
  \bibinfo {author} {\bibfnamefont {L.}~\bibnamefont {Lens}}, \bibinfo {author}
  {\bibfnamefont {E.}~\bibnamefont {Minaya~Ramirez}}, \bibinfo {author}
  {\bibfnamefont {A.~K.}\ \bibnamefont {Mistry}}, \bibinfo {author}
  {\bibfnamefont {S.}~\bibnamefont {Raeder}}, \bibinfo {author} {\bibfnamefont
  {P.}~\bibnamefont {Van~Duppen}}, \bibinfo {author} {\bibfnamefont
  {T.}~\bibnamefont {Walther}}, \bibinfo {author} {\bibfnamefont
  {A.}~\bibnamefont {Yakushev}}, \ and\ \bibinfo {author} {\bibfnamefont
  {Z.}~\bibnamefont {Zhang}},\ }\href {\doibase 10.1103/PhysRevLett.120.263003}
  {\bibfield  {journal} {\bibinfo  {journal} {Phys. Rev. Lett.}\ }\textbf
  {\bibinfo {volume} {120}},\ \bibinfo {pages} {263003} (\bibinfo {year}
  {2018})}\BibitemShut {NoStop}%
\bibitem [{\citenamefont {Ferrer}\ \emph {et~al.}(2013)\citenamefont {Ferrer},
  \citenamefont {Bastin}, \citenamefont {Boilley}, \citenamefont {Creemers},
  \citenamefont {Delahaye}, \citenamefont {Liénard}, \citenamefont
  {Fléchard}, \citenamefont {Franchoo}, \citenamefont {Ghys}, \citenamefont
  {Huyse}, \citenamefont {Kudryavtsev}, \citenamefont {Lecesne}, \citenamefont
  {Lu}, \citenamefont {Lutton}, \citenamefont {Mogilevskiy}, \citenamefont
  {Pauwels}, \citenamefont {Piot}, \citenamefont {Radulov}, \citenamefont
  {Rens}, \citenamefont {Savajols}, \citenamefont {Thomas}, \citenamefont
  {Traykov}, \citenamefont {{Van Beveren}}, \citenamefont {{Van den Bergh}},\
  and\ \citenamefont {{Van Duppen}}}]{Ferrer2013}%
  \BibitemOpen
  \bibfield  {author} {\bibinfo {author} {\bibfnamefont {R.}~\bibnamefont
  {Ferrer}}, \bibinfo {author} {\bibfnamefont {B.}~\bibnamefont {Bastin}},
  \bibinfo {author} {\bibfnamefont {D.}~\bibnamefont {Boilley}}, \bibinfo
  {author} {\bibfnamefont {P.}~\bibnamefont {Creemers}}, \bibinfo {author}
  {\bibfnamefont {P.}~\bibnamefont {Delahaye}}, \bibinfo {author}
  {\bibfnamefont {E.}~\bibnamefont {Liénard}}, \bibinfo {author}
  {\bibfnamefont {X.}~\bibnamefont {Fléchard}}, \bibinfo {author}
  {\bibfnamefont {S.}~\bibnamefont {Franchoo}}, \bibinfo {author}
  {\bibfnamefont {L.}~\bibnamefont {Ghys}}, \bibinfo {author} {\bibfnamefont
  {M.}~\bibnamefont {Huyse}}, \bibinfo {author} {\bibfnamefont
  {Y.}~\bibnamefont {Kudryavtsev}}, \bibinfo {author} {\bibfnamefont
  {N.}~\bibnamefont {Lecesne}}, \bibinfo {author} {\bibfnamefont
  {H.}~\bibnamefont {Lu}}, \bibinfo {author} {\bibfnamefont {F.}~\bibnamefont
  {Lutton}}, \bibinfo {author} {\bibfnamefont {E.}~\bibnamefont {Mogilevskiy}},
  \bibinfo {author} {\bibfnamefont {D.}~\bibnamefont {Pauwels}}, \bibinfo
  {author} {\bibfnamefont {J.}~\bibnamefont {Piot}}, \bibinfo {author}
  {\bibfnamefont {D.}~\bibnamefont {Radulov}}, \bibinfo {author} {\bibfnamefont
  {L.}~\bibnamefont {Rens}}, \bibinfo {author} {\bibfnamefont {H.}~\bibnamefont
  {Savajols}}, \bibinfo {author} {\bibfnamefont {J.}~\bibnamefont {Thomas}},
  \bibinfo {author} {\bibfnamefont {E.}~\bibnamefont {Traykov}}, \bibinfo
  {author} {\bibfnamefont {C.}~\bibnamefont {{Van Beveren}}}, \bibinfo {author}
  {\bibfnamefont {P.}~\bibnamefont {{Van den Bergh}}}, \ and\ \bibinfo {author}
  {\bibfnamefont {P.}~\bibnamefont {{Van Duppen}}},\ }\href {\doibase
  10.1016/j.nimb.2013.07.028} {\bibfield  {journal} {\bibinfo  {journal} {Nucl.
  Instrum. Meth. Phys. Res. Sect. B}\ }\textbf {\bibinfo {volume} {317}},\
  \bibinfo {pages} {570 } (\bibinfo {year} {2013})},\ \bibinfo {note} {xVIth
  International Conference on ElectroMagnetic Isotope Separators and Techniques
  Related to their Applications, December 2–7, 2012 at Matsue,
  Japan}\BibitemShut {NoStop}%
\bibitem [{\citenamefont {Ferrer}\ \emph {et~al.}(2017)\citenamefont {Ferrer},
  \citenamefont {Barzakh}, \citenamefont {Bastin}, \citenamefont {Beerwerth},
  \citenamefont {Block}, \citenamefont {Creemers}, \citenamefont {Grawe},
  \citenamefont {de~Groote}, \citenamefont {Delahaye}, \citenamefont
  {Fl{\'e}chard}, \citenamefont {Franchoo}, \citenamefont {Fritzsche},
  \citenamefont {Gaffney}, \citenamefont {Ghys}, \citenamefont {Gins},
  \citenamefont {Granados}, \citenamefont {Heinke}, \citenamefont {Hijazi},
  \citenamefont {Huyse}, \citenamefont {Kron}, \citenamefont {Kudryavtsev},
  \citenamefont {Laatiaoui}, \citenamefont {Lecesne}, \citenamefont {Loiselet},
  \citenamefont {Lutton}, \citenamefont {Moore}, \citenamefont {Martínez},
  \citenamefont {Mogilevskiy}, \citenamefont {Naubereit}, \citenamefont {Piot},
  \citenamefont {Raeder}, \citenamefont {Rothe}, \citenamefont {Savajols},
  \citenamefont {Sels}, \citenamefont {Sonnenschein}, \citenamefont {Thomas},
  \citenamefont {Traykov}, \citenamefont {Van~Beveren}, \citenamefont {Van~den
  Bergh}, \citenamefont {Van~Duppen}, \citenamefont {K.},\ and\ \citenamefont
  {Zadvornaya}}]{Ferrer2017towards}%
  \BibitemOpen
  \bibfield  {author} {\bibinfo {author} {\bibfnamefont {R.}~\bibnamefont
  {Ferrer}}, \bibinfo {author} {\bibfnamefont {A.}~\bibnamefont {Barzakh}},
  \bibinfo {author} {\bibfnamefont {B.}~\bibnamefont {Bastin}}, \bibinfo
  {author} {\bibfnamefont {R.}~\bibnamefont {Beerwerth}}, \bibinfo {author}
  {\bibfnamefont {M.}~\bibnamefont {Block}}, \bibinfo {author} {\bibfnamefont
  {P.}~\bibnamefont {Creemers}}, \bibinfo {author} {\bibfnamefont
  {H.}~\bibnamefont {Grawe}}, \bibinfo {author} {\bibfnamefont
  {R.}~\bibnamefont {de~Groote}}, \bibinfo {author} {\bibfnamefont
  {P.}~\bibnamefont {Delahaye}}, \bibinfo {author} {\bibfnamefont
  {X.}~\bibnamefont {Fl{\'e}chard}}, \bibinfo {author} {\bibfnamefont
  {S.}~\bibnamefont {Franchoo}}, \bibinfo {author} {\bibfnamefont
  {S.}~\bibnamefont {Fritzsche}}, \bibinfo {author} {\bibfnamefont {L.~P.}\
  \bibnamefont {Gaffney}}, \bibinfo {author} {\bibfnamefont {L.}~\bibnamefont
  {Ghys}}, \bibinfo {author} {\bibfnamefont {W.}~\bibnamefont {Gins}}, \bibinfo
  {author} {\bibfnamefont {C.}~\bibnamefont {Granados}}, \bibinfo {author}
  {\bibfnamefont {R.}~\bibnamefont {Heinke}}, \bibinfo {author} {\bibfnamefont
  {L.}~\bibnamefont {Hijazi}}, \bibinfo {author} {\bibfnamefont
  {M.}~\bibnamefont {Huyse}}, \bibinfo {author} {\bibfnamefont
  {T.}~\bibnamefont {Kron}}, \bibinfo {author} {\bibfnamefont {Y.}~\bibnamefont
  {Kudryavtsev}}, \bibinfo {author} {\bibfnamefont {M.}~\bibnamefont
  {Laatiaoui}}, \bibinfo {author} {\bibfnamefont {N.}~\bibnamefont {Lecesne}},
  \bibinfo {author} {\bibfnamefont {M.}~\bibnamefont {Loiselet}}, \bibinfo
  {author} {\bibfnamefont {F.}~\bibnamefont {Lutton}}, \bibinfo {author}
  {\bibfnamefont {I.}~\bibnamefont {Moore}}, \bibinfo {author} {\bibfnamefont
  {Y.}~\bibnamefont {Martínez}}, \bibinfo {author} {\bibfnamefont
  {E.}~\bibnamefont {Mogilevskiy}}, \bibinfo {author} {\bibfnamefont
  {P.}~\bibnamefont {Naubereit}}, \bibinfo {author} {\bibfnamefont
  {J.}~\bibnamefont {Piot}}, \bibinfo {author} {\bibfnamefont {S.}~\bibnamefont
  {Raeder}}, \bibinfo {author} {\bibfnamefont {S.}~\bibnamefont {Rothe}},
  \bibinfo {author} {\bibfnamefont {H.}~\bibnamefont {Savajols}}, \bibinfo
  {author} {\bibfnamefont {S.}~\bibnamefont {Sels}}, \bibinfo {author}
  {\bibfnamefont {V.}~\bibnamefont {Sonnenschein}}, \bibinfo {author}
  {\bibfnamefont {J.-C.}\ \bibnamefont {Thomas}}, \bibinfo {author}
  {\bibfnamefont {E.}~\bibnamefont {Traykov}}, \bibinfo {author} {\bibfnamefont
  {C.}~\bibnamefont {Van~Beveren}}, \bibinfo {author} {\bibfnamefont
  {P.}~\bibnamefont {Van~den Bergh}}, \bibinfo {author} {\bibfnamefont
  {P.}~\bibnamefont {Van~Duppen}}, \bibinfo {author} {\bibfnamefont
  {W.}~\bibnamefont {K.}}, \ and\ \bibinfo {author} {\bibfnamefont
  {A.}~\bibnamefont {Zadvornaya}},\ }\href {\doibase 10.1038/ncomms14520}
  {\bibfield  {journal} {\bibinfo  {journal} {Nature Commun.}\ }\textbf
  {\bibinfo {volume} {8}} (\bibinfo {year} {2017}),\
  10.1038/ncomms14520}\BibitemShut {NoStop}%
\bibitem [{\citenamefont {Sato}\ \emph {et~al.}(2015)\citenamefont {Sato},
  \citenamefont {Asai}, \citenamefont {Borschevsky}, \citenamefont {Stora},
  \citenamefont {Sato}, \citenamefont {Kaneya}, \citenamefont {Tsukada},
  \citenamefont {D{\"u}llmann}, \citenamefont {Eberhardt}, \citenamefont
  {Eliav} \emph {et~al.}}]{sato15a}%
  \BibitemOpen
  \bibfield  {author} {\bibinfo {author} {\bibfnamefont {T.}~\bibnamefont
  {Sato}}, \bibinfo {author} {\bibfnamefont {M.}~\bibnamefont {Asai}}, \bibinfo
  {author} {\bibfnamefont {A.}~\bibnamefont {Borschevsky}}, \bibinfo {author}
  {\bibfnamefont {T.}~\bibnamefont {Stora}}, \bibinfo {author} {\bibfnamefont
  {N.}~\bibnamefont {Sato}}, \bibinfo {author} {\bibfnamefont {Y.}~\bibnamefont
  {Kaneya}}, \bibinfo {author} {\bibfnamefont {K.}~\bibnamefont {Tsukada}},
  \bibinfo {author} {\bibfnamefont {C.~E.}\ \bibnamefont {D{\"u}llmann}},
  \bibinfo {author} {\bibfnamefont {K.}~\bibnamefont {Eberhardt}}, \bibinfo
  {author} {\bibfnamefont {E.}~\bibnamefont {Eliav}},  \emph {et~al.},\
  }\href@noop {} {\bibfield  {journal} {\bibinfo  {journal} {Nature}\ }\textbf
  {\bibinfo {volume} {520}},\ \bibinfo {pages} {209} (\bibinfo {year}
  {2015})}\BibitemShut {NoStop}%
\bibitem [{\citenamefont {Sato}\ \emph {et~al.}(2018)\citenamefont {Sato},
  \citenamefont {Asai}, \citenamefont {Borschevsky}, \citenamefont {Beerwerth},
  \citenamefont {Kaneya}, \citenamefont {Makii}, \citenamefont {Mitsukai},
  \citenamefont {Nagame}, \citenamefont {Osa}, \citenamefont {Toyoshima} \emph
  {et~al.}}]{sato18a}%
  \BibitemOpen
  \bibfield  {author} {\bibinfo {author} {\bibfnamefont {T.~K.}\ \bibnamefont
  {Sato}}, \bibinfo {author} {\bibfnamefont {M.}~\bibnamefont {Asai}}, \bibinfo
  {author} {\bibfnamefont {A.}~\bibnamefont {Borschevsky}}, \bibinfo {author}
  {\bibfnamefont {R.}~\bibnamefont {Beerwerth}}, \bibinfo {author}
  {\bibfnamefont {Y.}~\bibnamefont {Kaneya}}, \bibinfo {author} {\bibfnamefont
  {H.}~\bibnamefont {Makii}}, \bibinfo {author} {\bibfnamefont
  {A.}~\bibnamefont {Mitsukai}}, \bibinfo {author} {\bibfnamefont
  {Y.}~\bibnamefont {Nagame}}, \bibinfo {author} {\bibfnamefont
  {A.}~\bibnamefont {Osa}}, \bibinfo {author} {\bibfnamefont {A.}~\bibnamefont
  {Toyoshima}},  \emph {et~al.},\ }\href@noop {} {\bibfield  {journal}
  {\bibinfo  {journal} {Journal of the American Chemical Society}\ }\textbf
  {\bibinfo {volume} {140}},\ \bibinfo {pages} {14609} (\bibinfo {year}
  {2018})}\BibitemShut {NoStop}%
\bibitem [{\citenamefont {Lautenschl{\"a}ger}\ \emph
  {et~al.}(2016)\citenamefont {Lautenschl{\"a}ger}, \citenamefont {Chhetri},
  \citenamefont {Ackermann}, \citenamefont {Backe}, \citenamefont {Block},
  \citenamefont {Cheal}, \citenamefont {Clark}, \citenamefont {Droese},
  \citenamefont {Ferrer}, \citenamefont {Giacoppo}, \citenamefont {G{\"o}tz},
  \citenamefont {He{\ss}berger}, \citenamefont {Kaleja}, \citenamefont
  {Khuyagbaatar}, \citenamefont {Kunz}, \citenamefont {Mistry}, \citenamefont
  {Laatiaoui}, \citenamefont {Lauth}, \citenamefont {Raeder}, \citenamefont
  {Walther},\ and\ \citenamefont {Wraith}}]{Lautenschlaeger2016115}%
  \BibitemOpen
  \bibfield  {author} {\bibinfo {author} {\bibfnamefont {F.}~\bibnamefont
  {Lautenschl{\"a}ger}}, \bibinfo {author} {\bibfnamefont {P.}~\bibnamefont
  {Chhetri}}, \bibinfo {author} {\bibfnamefont {D.}~\bibnamefont {Ackermann}},
  \bibinfo {author} {\bibfnamefont {H.}~\bibnamefont {Backe}}, \bibinfo
  {author} {\bibfnamefont {M.}~\bibnamefont {Block}}, \bibinfo {author}
  {\bibfnamefont {B.}~\bibnamefont {Cheal}}, \bibinfo {author} {\bibfnamefont
  {A.}~\bibnamefont {Clark}}, \bibinfo {author} {\bibfnamefont
  {C.}~\bibnamefont {Droese}}, \bibinfo {author} {\bibfnamefont
  {R.}~\bibnamefont {Ferrer}}, \bibinfo {author} {\bibfnamefont
  {F.}~\bibnamefont {Giacoppo}}, \bibinfo {author} {\bibfnamefont
  {S.}~\bibnamefont {G{\"o}tz}}, \bibinfo {author} {\bibfnamefont
  {F.}~\bibnamefont {He{\ss}berger}}, \bibinfo {author} {\bibfnamefont
  {O.}~\bibnamefont {Kaleja}}, \bibinfo {author} {\bibfnamefont
  {J.}~\bibnamefont {Khuyagbaatar}}, \bibinfo {author} {\bibfnamefont
  {P.}~\bibnamefont {Kunz}}, \bibinfo {author} {\bibfnamefont {A.}~\bibnamefont
  {Mistry}}, \bibinfo {author} {\bibfnamefont {M.}~\bibnamefont {Laatiaoui}},
  \bibinfo {author} {\bibfnamefont {W.}~\bibnamefont {Lauth}}, \bibinfo
  {author} {\bibfnamefont {S.}~\bibnamefont {Raeder}}, \bibinfo {author}
  {\bibfnamefont {T.}~\bibnamefont {Walther}}, \ and\ \bibinfo {author}
  {\bibfnamefont {C.}~\bibnamefont {Wraith}},\ }\href {\doibase
  10.1016/j.nimb.2016.06.001} {\bibfield  {journal} {\bibinfo  {journal} {Nucl.
  Instrum. Meth. Phys. Res. B}\ }\textbf {\bibinfo {volume} {383}},\ \bibinfo
  {pages} {115} (\bibinfo {year} {2016})}\BibitemShut {NoStop}%
\bibitem [{\citenamefont {Murb{\"o}ck}\ \emph {et~al.}(2020)\citenamefont
  {Murb{\"o}ck}, \citenamefont {Raeder}, \citenamefont {Chhetri}, \citenamefont
  {Diaz}, \citenamefont {Laatiaoui}, \citenamefont {Giacoppo},\ and\
  \citenamefont {Block}}]{Murboeck2020}%
  \BibitemOpen
  \bibfield  {author} {\bibinfo {author} {\bibfnamefont {T.}~\bibnamefont
  {Murb{\"o}ck}}, \bibinfo {author} {\bibfnamefont {S.}~\bibnamefont {Raeder}},
  \bibinfo {author} {\bibfnamefont {P.}~\bibnamefont {Chhetri}}, \bibinfo
  {author} {\bibfnamefont {K.}~\bibnamefont {Diaz}}, \bibinfo {author}
  {\bibfnamefont {M.}~\bibnamefont {Laatiaoui}}, \bibinfo {author}
  {\bibfnamefont {F.}~\bibnamefont {Giacoppo}}, \ and\ \bibinfo {author}
  {\bibfnamefont {M.}~\bibnamefont {Block}},\ }\href {\doibase
  10.1007/s10751-019-1689-1} {\bibfield  {journal} {\bibinfo  {journal}
  {Hyperfine Interact.}\ }\textbf {\bibinfo {volume} {241}},\ \bibinfo {pages}
  {35} (\bibinfo {year} {2020})}\BibitemShut {NoStop}%
\bibitem [{\citenamefont {Gäggeler}\ \emph {et~al.}(1989)\citenamefont
  {Gäggeler}, \citenamefont {Jost}, \citenamefont {Türler}, \citenamefont
  {Armbruster}, \citenamefont {Brüchle}, \citenamefont {Folger}, \citenamefont
  {He{\ss}berger}, \citenamefont {Hofmann}, \citenamefont {Münzenberg},
  \citenamefont {Ninov}, \citenamefont {Reisdorf}, \citenamefont {Schädel},
  \citenamefont {Sümmerer}, \citenamefont {Kratz}, \citenamefont {Scherer},\
  and\ \citenamefont {Leino}}]{Gaeggeler1989}%
  \BibitemOpen
  \bibfield  {author} {\bibinfo {author} {\bibfnamefont {H.}~\bibnamefont
  {Gäggeler}}, \bibinfo {author} {\bibfnamefont {D.}~\bibnamefont {Jost}},
  \bibinfo {author} {\bibfnamefont {A.}~\bibnamefont {Türler}}, \bibinfo
  {author} {\bibfnamefont {P.}~\bibnamefont {Armbruster}}, \bibinfo {author}
  {\bibfnamefont {W.}~\bibnamefont {Brüchle}}, \bibinfo {author}
  {\bibfnamefont {H.}~\bibnamefont {Folger}}, \bibinfo {author} {\bibfnamefont
  {F.}~\bibnamefont {He{\ss}berger}}, \bibinfo {author} {\bibfnamefont
  {S.}~\bibnamefont {Hofmann}}, \bibinfo {author} {\bibfnamefont
  {G.}~\bibnamefont {Münzenberg}}, \bibinfo {author} {\bibfnamefont
  {V.}~\bibnamefont {Ninov}}, \bibinfo {author} {\bibfnamefont
  {W.}~\bibnamefont {Reisdorf}}, \bibinfo {author} {\bibfnamefont
  {M.}~\bibnamefont {Schädel}}, \bibinfo {author} {\bibfnamefont
  {K.}~\bibnamefont {Sümmerer}}, \bibinfo {author} {\bibfnamefont
  {J.}~\bibnamefont {Kratz}}, \bibinfo {author} {\bibfnamefont
  {U.}~\bibnamefont {Scherer}}, \ and\ \bibinfo {author} {\bibfnamefont
  {M.}~\bibnamefont {Leino}},\ }\href {\doibase 10.1016/0375-9474(89)90689-1}
  {\bibfield  {journal} {\bibinfo  {journal} {Nuclear Physics A}\ }\textbf
  {\bibinfo {volume} {502}},\ \bibinfo {pages} {561} (\bibinfo {year}
  {1989})}\BibitemShut {NoStop}%
\bibitem [{\citenamefont {Borschevsky}\ \emph {et~al.}(2007)\citenamefont
  {Borschevsky}, \citenamefont {Eliav}, \citenamefont {Vilkas}, \citenamefont
  {Ishikawa},\ and\ \citenamefont {Kaldor}}]{borschevsky07a}%
  \BibitemOpen
  \bibfield  {author} {\bibinfo {author} {\bibfnamefont {A.}~\bibnamefont
  {Borschevsky}}, \bibinfo {author} {\bibfnamefont {E.}~\bibnamefont {Eliav}},
  \bibinfo {author} {\bibfnamefont {M.}~\bibnamefont {Vilkas}}, \bibinfo
  {author} {\bibfnamefont {Y.}~\bibnamefont {Ishikawa}}, \ and\ \bibinfo
  {author} {\bibfnamefont {U.}~\bibnamefont {Kaldor}},\ }\href@noop {}
  {\bibfield  {journal} {\bibinfo  {journal} {The European Physical Journal D}\
  }\textbf {\bibinfo {volume} {45}},\ \bibinfo {pages} {115} (\bibinfo {year}
  {2007})}\BibitemShut {NoStop}%
\bibitem [{\citenamefont {Dzuba}\ \emph {et~al.}(2014)\citenamefont {Dzuba},
  \citenamefont {Safronova},\ and\ \citenamefont {Safronova}}]{dzuba14a}%
  \BibitemOpen
  \bibfield  {author} {\bibinfo {author} {\bibfnamefont {V.~A.}\ \bibnamefont
  {Dzuba}}, \bibinfo {author} {\bibfnamefont {M.~S.}\ \bibnamefont
  {Safronova}}, \ and\ \bibinfo {author} {\bibfnamefont {U.~I.}\ \bibnamefont
  {Safronova}},\ }\href {\doibase 10.1103/PhysRevA.90.012504} {\bibfield
  {journal} {\bibinfo  {journal} {Phys. Rev. A}\ }\textbf {\bibinfo {volume}
  {90}},\ \bibinfo {pages} {012504} (\bibinfo {year} {2014})}\BibitemShut
  {NoStop}%
\bibitem [{\citenamefont {Fritzsche}\ \emph {et~al.}(2007)\citenamefont
  {Fritzsche}, \citenamefont {Dong}, \citenamefont {Koike},\ and\ \citenamefont
  {Uvarov}}]{fritzsche07a}%
  \BibitemOpen
  \bibfield  {author} {\bibinfo {author} {\bibfnamefont {S.}~\bibnamefont
  {Fritzsche}}, \bibinfo {author} {\bibfnamefont {C.}~\bibnamefont {Dong}},
  \bibinfo {author} {\bibfnamefont {F.}~\bibnamefont {Koike}}, \ and\ \bibinfo
  {author} {\bibfnamefont {A.}~\bibnamefont {Uvarov}},\ }\href@noop {}
  {\bibfield  {journal} {\bibinfo  {journal} {The European Physical Journal D}\
  }\textbf {\bibinfo {volume} {45}},\ \bibinfo {pages} {107} (\bibinfo {year}
  {2007})}\BibitemShut {NoStop}%
\bibitem [{\citenamefont {Zou}\ and\ \citenamefont
  {Froese~Fischer}(2002)}]{zou02a}%
  \BibitemOpen
  \bibfield  {author} {\bibinfo {author} {\bibfnamefont {Y.}~\bibnamefont
  {Zou}}\ and\ \bibinfo {author} {\bibfnamefont {C.}~\bibnamefont
  {Froese~Fischer}},\ }\href {\doibase 10.1103/PhysRevLett.88.183001}
  {\bibfield  {journal} {\bibinfo  {journal} {Phys. Rev. Lett.}\ }\textbf
  {\bibinfo {volume} {88}},\ \bibinfo {pages} {183001} (\bibinfo {year}
  {2002})}\BibitemShut {NoStop}%
\bibitem [{\citenamefont {Pa\ifmmode~\check{s}\else \v{s}\fi{}teka}\ \emph
  {et~al.}(2017)\citenamefont {Pa\ifmmode~\check{s}\else \v{s}\fi{}teka},
  \citenamefont {Eliav}, \citenamefont {Borschevsky}, \citenamefont {Kaldor},\
  and\ \citenamefont {Schwerdtfeger}}]{PasEliBor17}%
  \BibitemOpen
  \bibfield  {author} {\bibinfo {author} {\bibfnamefont {L.~F.}\ \bibnamefont
  {Pa\ifmmode~\check{s}\else \v{s}\fi{}teka}}, \bibinfo {author} {\bibfnamefont
  {E.}~\bibnamefont {Eliav}}, \bibinfo {author} {\bibfnamefont
  {A.}~\bibnamefont {Borschevsky}}, \bibinfo {author} {\bibfnamefont
  {U.}~\bibnamefont {Kaldor}}, \ and\ \bibinfo {author} {\bibfnamefont
  {P.}~\bibnamefont {Schwerdtfeger}},\ }\href {\doibase
  10.1103/PhysRevLett.118.023002} {\bibfield  {journal} {\bibinfo  {journal}
  {Phys. Rev. Lett.}\ }\textbf {\bibinfo {volume} {118}},\ \bibinfo {pages}
  {023002} (\bibinfo {year} {2017})}\BibitemShut {NoStop}%
\bibitem [{\citenamefont {Kahl}\ \emph {et~al.}(2019)\citenamefont {Kahl},
  \citenamefont {Berengut}, \citenamefont {Laatiaoui}, \citenamefont {Eliav},\
  and\ \citenamefont {Borschevsky}}]{kahl19b}%
  \BibitemOpen
  \bibfield  {author} {\bibinfo {author} {\bibfnamefont {E.~V.}\ \bibnamefont
  {Kahl}}, \bibinfo {author} {\bibfnamefont {J.~C.}\ \bibnamefont {Berengut}},
  \bibinfo {author} {\bibfnamefont {M.}~\bibnamefont {Laatiaoui}}, \bibinfo
  {author} {\bibfnamefont {E.}~\bibnamefont {Eliav}}, \ and\ \bibinfo {author}
  {\bibfnamefont {A.}~\bibnamefont {Borschevsky}},\ }\href {\doibase
  10.1103/PhysRevA.100.062505} {\bibfield  {journal} {\bibinfo  {journal}
  {Phys. Rev. A}\ }\textbf {\bibinfo {volume} {100}},\ \bibinfo {pages}
  {062505} (\bibinfo {year} {2019})}\BibitemShut {NoStop}%
\bibitem [{\citenamefont {Sucher}(1980)}]{Suc80}%
  \BibitemOpen
  \bibfield  {author} {\bibinfo {author} {\bibfnamefont {J.}~\bibnamefont
  {Sucher}},\ }\href {\doibase 10.1103/PhysRevA.22.348} {\bibfield  {journal}
  {\bibinfo  {journal} {Phys. Rev. A}\ }\textbf {\bibinfo {volume} {22}},\
  \bibinfo {pages} {348} (\bibinfo {year} {1980})}\BibitemShut {NoStop}%
\bibitem [{\citenamefont {Visscher}\ and\ \citenamefont
  {Dyall}(1997)}]{VisDya97}%
  \BibitemOpen
  \bibfield  {author} {\bibinfo {author} {\bibfnamefont {L.}~\bibnamefont
  {Visscher}}\ and\ \bibinfo {author} {\bibfnamefont {K.~G.}\ \bibnamefont
  {Dyall}},\ }\href {\doibase 10.1006/adnd.1997.0751} {\bibfield  {journal}
  {\bibinfo  {journal} {At. Data Nucl. Data Tabl.}\ }\textbf {\bibinfo {volume}
  {67}},\ \bibinfo {pages} {207 } (\bibinfo {year} {1997})}\BibitemShut
  {NoStop}%
\bibitem [{DIR()}]{DIRAC15}%
  \BibitemOpen
  \href@noop {} {}\bibinfo {note} {DIRAC, a relativistic ab initio electronic
  structure program, Release DIRAC15 (2015), written by R. Bast, T. Saue, L.
  Visscher, and H. J. Aa. Jensen, with contributions from V. Bakken, K. G.
  Dyall, S. Dubillard, U. Ekstroem, E. Eliav, T. Enevoldsen, E. Fasshauer, T.
  Fleig, O. Fossgaard, A. S. P. Gomes, T. Helgaker, J. Henriksson, M. Ilias,
  Ch. R. Jacob, S. Knecht, S. Komorovsky, O. Kullie, J. K. Laerdahl, C. V.
  Larsen, Y. S. Lee, H. S. Nataraj, M. K. Nayak, P. Norman, G. Olejniczak, J.
  Olsen, Y. C. Park, J. K. Pedersen, M. Pernpointner, R. Di Remigio, K. Ruud,
  P. Salek, B. Schimmelpfennig, J. Sikkema, A. J. Thorvaldsen, J. Thyssen, J.
  van Stralen, S. Villaume, O. Visser, T. Winther, and S. Yamamoto (see
  http://www.diracprogram.org).}\BibitemShut {Stop}%
\bibitem [{\citenamefont {Dyall}(2006)}]{Dya06}%
  \BibitemOpen
  \bibfield  {author} {\bibinfo {author} {\bibfnamefont {K.~G.}\ \bibnamefont
  {Dyall}},\ }\href {\doibase 10.1007/s00214-006-0126-0} {\bibfield  {journal}
  {\bibinfo  {journal} {Theoretical Chemistry Accounts}\ }\textbf {\bibinfo
  {volume} {115}},\ \bibinfo {pages} {441} (\bibinfo {year}
  {2006})}\BibitemShut {NoStop}%
\bibitem [{\citenamefont {Helgaker}\ \emph {et~al.}(1997)\citenamefont
  {Helgaker}, \citenamefont {Klopper}, \citenamefont {Koch},\ and\
  \citenamefont {Noga}}]{HelKloKoc97}%
  \BibitemOpen
  \bibfield  {author} {\bibinfo {author} {\bibfnamefont {T.}~\bibnamefont
  {Helgaker}}, \bibinfo {author} {\bibfnamefont {W.}~\bibnamefont {Klopper}},
  \bibinfo {author} {\bibfnamefont {H.}~\bibnamefont {Koch}}, \ and\ \bibinfo
  {author} {\bibfnamefont {J.}~\bibnamefont {Noga}},\ }\href {\doibase
  10.1063/1.473863} {\bibfield  {journal} {\bibinfo  {journal} {The Journal of
  Chemical Physics}\ }\textbf {\bibinfo {volume} {106}},\ \bibinfo {pages}
  {9639} (\bibinfo {year} {1997})}\BibitemShut {NoStop}%
\bibitem [{TRA()}]{TRAFS-3C}%
  \BibitemOpen
  \href@noop {} {}\bibinfo {note} {TRAFS-3C code (Tel-Aviv Relativistic Atomic
  Fock-Space coupled cluster code), written by E.Eliav, U.Kaldor and Y.Ishikawa
  (1990-2013), with contributions by A. Landau}\BibitemShut {NoStop}%
\bibitem [{\citenamefont {Dzuba}\ \emph {et~al.}(1996)\citenamefont {Dzuba},
  \citenamefont {Flambaum},\ and\ \citenamefont {Kozlov}}]{dzuba96a}%
  \BibitemOpen
  \bibfield  {author} {\bibinfo {author} {\bibfnamefont {V.~A.}\ \bibnamefont
  {Dzuba}}, \bibinfo {author} {\bibfnamefont {V.~V.}\ \bibnamefont {Flambaum}},
  \ and\ \bibinfo {author} {\bibfnamefont {M.~G.}\ \bibnamefont {Kozlov}},\
  }\href@noop {} {\bibfield  {journal} {\bibinfo  {journal} {Phys.\ Rev.\ A}\
  }\textbf {\bibinfo {volume} {54}},\ \bibinfo {pages} {3948} (\bibinfo {year}
  {1996})}\BibitemShut {NoStop}%
\bibitem [{\citenamefont {Kahl}\ and\ \citenamefont
  {Berengut}(2019)}]{kahl19a}%
  \BibitemOpen
  \bibfield  {author} {\bibinfo {author} {\bibfnamefont {E.~V.}\ \bibnamefont
  {Kahl}}\ and\ \bibinfo {author} {\bibfnamefont {J.~C.}\ \bibnamefont
  {Berengut}},\ }\href {\doibase https://doi.org/10.1016/j.cpc.2018.12.014}
  {\bibfield  {journal} {\bibinfo  {journal} {Computer Physics Communications}\
  }\textbf {\bibinfo {volume} {238}},\ \bibinfo {pages} {232 } (\bibinfo {year}
  {2019})}\BibitemShut {NoStop}%
\bibitem [{\citenamefont {Berengut}\ \emph {et~al.}(2006)\citenamefont
  {Berengut}, \citenamefont {Flambaum},\ and\ \citenamefont
  {Kozlov}}]{berengut06a}%
  \BibitemOpen
  \bibfield  {author} {\bibinfo {author} {\bibfnamefont {J.~C.}\ \bibnamefont
  {Berengut}}, \bibinfo {author} {\bibfnamefont {V.~V.}\ \bibnamefont
  {Flambaum}}, \ and\ \bibinfo {author} {\bibfnamefont {M.~G.}\ \bibnamefont
  {Kozlov}},\ }\href@noop {} {\bibfield  {journal} {\bibinfo  {journal} {Phys.
  Rev. A}\ }\textbf {\bibinfo {volume} {73}},\ \bibinfo {pages} {012504}
  (\bibinfo {year} {2006})}\BibitemShut {NoStop}%
\bibitem [{\citenamefont {Berengut}(2016)}]{berengut16a}%
  \BibitemOpen
  \bibfield  {author} {\bibinfo {author} {\bibfnamefont {J.~C.}\ \bibnamefont
  {Berengut}},\ }\href@noop {} {\bibfield  {journal} {\bibinfo  {journal}
  {Phys. Rev. A}\ }\textbf {\bibinfo {volume} {94}},\ \bibinfo {pages} {012502}
  (\bibinfo {year} {2016})}\BibitemShut {NoStop}%
\bibitem [{\citenamefont {Geddes}\ \emph {et~al.}(2018)\citenamefont {Geddes},
  \citenamefont {Czapski}, \citenamefont {Kahl},\ and\ \citenamefont
  {Berengut}}]{geddes18a}%
  \BibitemOpen
  \bibfield  {author} {\bibinfo {author} {\bibfnamefont {A.~J.}\ \bibnamefont
  {Geddes}}, \bibinfo {author} {\bibfnamefont {D.~A.}\ \bibnamefont {Czapski}},
  \bibinfo {author} {\bibfnamefont {E.~V.}\ \bibnamefont {Kahl}}, \ and\
  \bibinfo {author} {\bibfnamefont {J.~C.}\ \bibnamefont {Berengut}},\ }\href
  {\doibase 10.1103/PhysRevA.98.042508} {\bibfield  {journal} {\bibinfo
  {journal} {Phys. Rev. A}\ }\textbf {\bibinfo {volume} {98}},\ \bibinfo
  {pages} {042508} (\bibinfo {year} {2018})}\BibitemShut {NoStop}%
\bibitem [{\citenamefont {Flambaum}\ and\ \citenamefont
  {Ginges}(2005)}]{flambaum05a}%
  \BibitemOpen
  \bibfield  {author} {\bibinfo {author} {\bibfnamefont {V.~V.}\ \bibnamefont
  {Flambaum}}\ and\ \bibinfo {author} {\bibfnamefont {J.~S.~M.}\ \bibnamefont
  {Ginges}},\ }\href@noop {} {\bibfield  {journal} {\bibinfo  {journal} {Phys.
  Rev. A}\ }\textbf {\bibinfo {volume} {72}},\ \bibinfo {pages} {052115}
  (\bibinfo {year} {2005})}\BibitemShut {NoStop}%
\bibitem [{\citenamefont {Ginges}\ and\ \citenamefont
  {Berengut}(2016{\natexlab{a}})}]{ginges16a}%
  \BibitemOpen
  \bibfield  {author} {\bibinfo {author} {\bibfnamefont {J.~S.~M.}\
  \bibnamefont {Ginges}}\ and\ \bibinfo {author} {\bibfnamefont {J.~C.}\
  \bibnamefont {Berengut}},\ }\href@noop {} {\bibfield  {journal} {\bibinfo
  {journal} {Phys. Rev. A}\ }\textbf {\bibinfo {volume} {93}},\ \bibinfo
  {pages} {052509} (\bibinfo {year} {2016}{\natexlab{a}})}\BibitemShut
  {NoStop}%
\bibitem [{\citenamefont {Ginges}\ and\ \citenamefont
  {Berengut}(2016{\natexlab{b}})}]{ginges16b}%
  \BibitemOpen
  \bibfield  {author} {\bibinfo {author} {\bibfnamefont {J.~S.~M.}\
  \bibnamefont {Ginges}}\ and\ \bibinfo {author} {\bibfnamefont {J.~C.}\
  \bibnamefont {Berengut}},\ }\href@noop {} {\bibfield  {journal} {\bibinfo
  {journal} {J. Phys. B}\ }\textbf {\bibinfo {volume} {49}},\ \bibinfo {pages}
  {095001} (\bibinfo {year} {2016}{\natexlab{b}})}\BibitemShut {NoStop}%
\bibitem [{\citenamefont {Johnson}\ \emph {et~al.}(1988)\citenamefont
  {Johnson}, \citenamefont {Blundell},\ and\ \citenamefont
  {Sapirstein}}]{johnson88a}%
  \BibitemOpen
  \bibfield  {author} {\bibinfo {author} {\bibfnamefont {W.~R.}\ \bibnamefont
  {Johnson}}, \bibinfo {author} {\bibfnamefont {S.~A.}\ \bibnamefont
  {Blundell}}, \ and\ \bibinfo {author} {\bibfnamefont {J.}~\bibnamefont
  {Sapirstein}},\ }\href@noop {} {\bibfield  {journal} {\bibinfo  {journal}
  {Phys.~Rev.~A}\ }\textbf {\bibinfo {volume} {37}},\ \bibinfo {pages} {307}
  (\bibinfo {year} {1988})}\BibitemShut {NoStop}%
\bibitem [{\citenamefont {{Beloy}}\ and\ \citenamefont
  {{Derevianko}}(2008)}]{beloy08a}%
  \BibitemOpen
  \bibfield  {author} {\bibinfo {author} {\bibfnamefont {K.}~\bibnamefont
  {{Beloy}}}\ and\ \bibinfo {author} {\bibfnamefont {A.}~\bibnamefont
  {{Derevianko}}},\ }\href@noop {} {\bibfield  {journal} {\bibinfo  {journal}
  {Comp.~Phys.~Commun.}\ }\textbf {\bibinfo {volume} {179}},\ \bibinfo {pages}
  {310} (\bibinfo {year} {2008})}\BibitemShut {NoStop}%
\bibitem [{\citenamefont {Shabaev}\ \emph {et~al.}(2004)\citenamefont
  {Shabaev}, \citenamefont {Tupitsyn}, \citenamefont {Yerokhin}, \citenamefont
  {Plunien},\ and\ \citenamefont {Soff}}]{shabaev04a}%
  \BibitemOpen
  \bibfield  {author} {\bibinfo {author} {\bibfnamefont {V.}~\bibnamefont
  {Shabaev}}, \bibinfo {author} {\bibfnamefont {I.}~\bibnamefont {Tupitsyn}},
  \bibinfo {author} {\bibfnamefont {V.}~\bibnamefont {Yerokhin}}, \bibinfo
  {author} {\bibfnamefont {G.}~\bibnamefont {Plunien}}, \ and\ \bibinfo
  {author} {\bibfnamefont {G.}~\bibnamefont {Soff}},\ }\href@noop {} {\bibfield
   {journal} {\bibinfo  {journal} {Physical review letters}\ }\textbf {\bibinfo
  {volume} {93}},\ \bibinfo {pages} {130405} (\bibinfo {year}
  {2004})}\BibitemShut {NoStop}%
\bibitem [{\citenamefont {Dzuba}\ \emph
  {et~al.}(2017{\natexlab{a}})\citenamefont {Dzuba}, \citenamefont {Berengut},
  \citenamefont {Harabati},\ and\ \citenamefont {Flambaum}}]{dzuba17a}%
  \BibitemOpen
  \bibfield  {author} {\bibinfo {author} {\bibfnamefont {V.~A.}\ \bibnamefont
  {Dzuba}}, \bibinfo {author} {\bibfnamefont {J.~C.}\ \bibnamefont {Berengut}},
  \bibinfo {author} {\bibfnamefont {C.}~\bibnamefont {Harabati}}, \ and\
  \bibinfo {author} {\bibfnamefont {V.~V.}\ \bibnamefont {Flambaum}},\
  }\href@noop {} {\bibfield  {journal} {\bibinfo  {journal} {Physical Review
  A}\ }\textbf {\bibinfo {volume} {95}},\ \bibinfo {pages} {012503} (\bibinfo
  {year} {2017}{\natexlab{a}})}\BibitemShut {NoStop}%
\bibitem [{\citenamefont {Dzuba}(2005)}]{dzuba05a}%
  \BibitemOpen
  \bibfield  {author} {\bibinfo {author} {\bibfnamefont {V.~A.}\ \bibnamefont
  {Dzuba}},\ }\href {\doibase 10.1103/PhysRevA.71.032512} {\bibfield  {journal}
  {\bibinfo  {journal} {Phys. Rev. A}\ }\textbf {\bibinfo {volume} {71}},\
  \bibinfo {pages} {032512} (\bibinfo {year} {2005})}\BibitemShut {NoStop}%
\bibitem [{\citenamefont {Dzuba}\ \emph {et~al.}(1998)\citenamefont {Dzuba},
  \citenamefont {Flambaum}, \citenamefont {Kozlov},\ and\ \citenamefont
  {Porsev}}]{dzuba98jetp}%
  \BibitemOpen
  \bibfield  {author} {\bibinfo {author} {\bibfnamefont {V.~A.}\ \bibnamefont
  {Dzuba}}, \bibinfo {author} {\bibfnamefont {V.~V.}\ \bibnamefont {Flambaum}},
  \bibinfo {author} {\bibfnamefont {M.~G.}\ \bibnamefont {Kozlov}}, \ and\
  \bibinfo {author} {\bibfnamefont {S.~G.}\ \bibnamefont {Porsev}},\
  }\href@noop {} {\bibfield  {journal} {\bibinfo  {journal} {Journal of
  Experimental and Theoretical Physics}\ }\textbf {\bibinfo {volume} {87}},\
  \bibinfo {pages} {885} (\bibinfo {year} {1998})}\BibitemShut {NoStop}%
\bibitem [{\citenamefont {Porsev}\ \emph {et~al.}(1999)\citenamefont {Porsev},
  \citenamefont {Rakhlina},\ and\ \citenamefont {Kozlov}}]{porsev99pra}%
  \BibitemOpen
  \bibfield  {author} {\bibinfo {author} {\bibfnamefont {S.~G.}\ \bibnamefont
  {Porsev}}, \bibinfo {author} {\bibfnamefont {Y.~G.}\ \bibnamefont
  {Rakhlina}}, \ and\ \bibinfo {author} {\bibfnamefont {M.~G.}\ \bibnamefont
  {Kozlov}},\ }\href@noop {} {\bibfield  {journal} {\bibinfo  {journal}
  {Physical Review A}\ }\textbf {\bibinfo {volume} {60}},\ \bibinfo {pages}
  {2781} (\bibinfo {year} {1999})}\BibitemShut {NoStop}%
\bibitem [{\citenamefont {Porsev}\ \emph {et~al.}(2001)\citenamefont {Porsev},
  \citenamefont {Kozlov}, \citenamefont {Rakhlina},\ and\ \citenamefont
  {Derevianko}}]{porsev01pra}%
  \BibitemOpen
  \bibfield  {author} {\bibinfo {author} {\bibfnamefont {S.~G.}\ \bibnamefont
  {Porsev}}, \bibinfo {author} {\bibfnamefont {M.~G.}\ \bibnamefont {Kozlov}},
  \bibinfo {author} {\bibfnamefont {Y.~G.}\ \bibnamefont {Rakhlina}}, \ and\
  \bibinfo {author} {\bibfnamefont {A.}~\bibnamefont {Derevianko}},\
  }\href@noop {} {\bibfield  {journal} {\bibinfo  {journal} {Physical Review
  A}\ }\textbf {\bibinfo {volume} {64}},\ \bibinfo {pages} {012508} (\bibinfo
  {year} {2001})}\BibitemShut {NoStop}%
\bibitem [{\citenamefont {Kramida}\ \emph {et~al.}(2016)\citenamefont
  {Kramida}, \citenamefont {Ralchenko}, \citenamefont {Reader},\ and\
  \citenamefont {{NIST ASD Team}}}]{nistasd54}%
  \BibitemOpen
  \bibfield  {author} {\bibinfo {author} {\bibfnamefont {A.}~\bibnamefont
  {Kramida}}, \bibinfo {author} {\bibfnamefont {Y.}~\bibnamefont {Ralchenko}},
  \bibinfo {author} {\bibfnamefont {J.}~\bibnamefont {Reader}}, \ and\ \bibinfo
  {author} {\bibnamefont {{NIST ASD Team}}},\ }\href {\doibase 10.18434/T4W30F}
  {\enquote {\bibinfo {title} {{NIST Atomic Spectra Database (v5.4)}},}\ }
  (\bibinfo {year} {2016})\BibitemShut {NoStop}%
\bibitem [{\citenamefont {{Martin}}\ \emph {et~al.}(1978)\citenamefont
  {{Martin}}, \citenamefont {{Zalubas}},\ and\ \citenamefont
  {{Hagan}}}]{MarZalHag78}%
  \BibitemOpen
  \bibfield  {author} {\bibinfo {author} {\bibfnamefont {W.~C.}\ \bibnamefont
  {{Martin}}}, \bibinfo {author} {\bibfnamefont {R.}~\bibnamefont {{Zalubas}}},
  \ and\ \bibinfo {author} {\bibfnamefont {L.}~\bibnamefont {{Hagan}}},\
  }\href@noop {} {\emph {\bibinfo {title} {NSRDS-NBS, Washington: National
  Bureau of Standards, U.S.~Department of Commerce, |c1978}}}\ (\bibinfo {year}
  {1978})\BibitemShut {NoStop}%
\bibitem [{\citenamefont {Sobel'man}(1972)}]{sobelman72book}%
  \BibitemOpen
  \bibfield  {author} {\bibinfo {author} {\bibfnamefont {I.~I.}\ \bibnamefont
  {Sobel'man}},\ }\href@noop {} {\emph {\bibinfo {title} {Introduction to the
  Theory of Atomic Spectra}}}\ (\bibinfo  {publisher} {Pergamon},\ \bibinfo
  {address} {New York},\ \bibinfo {year} {1972})\BibitemShut {NoStop}%
\bibitem [{\citenamefont {Dzuba}\ \emph
  {et~al.}(2017{\natexlab{b}})\citenamefont {Dzuba}, \citenamefont {Flambaum},\
  and\ \citenamefont {Webb}}]{dzuba17pra0}%
  \BibitemOpen
  \bibfield  {author} {\bibinfo {author} {\bibfnamefont {V.~A.}\ \bibnamefont
  {Dzuba}}, \bibinfo {author} {\bibfnamefont {V.~V.}\ \bibnamefont {Flambaum}},
  \ and\ \bibinfo {author} {\bibfnamefont {J.~K.}\ \bibnamefont {Webb}},\
  }\href@noop {} {\bibfield  {journal} {\bibinfo  {journal} {\pra}\ }\textbf
  {\bibinfo {volume} {95}},\ \bibinfo {pages} {062515} (\bibinfo {year}
  {2017}{\natexlab{b}})}\BibitemShut {NoStop}%
\bibitem [{\citenamefont {Chhetri}\ \emph {et~al.}(2017)\citenamefont
  {Chhetri}, \citenamefont {Ackermann}, \citenamefont {Backe}, \citenamefont
  {Block}, \citenamefont {Cheal}, \citenamefont {D{\"u}llmann}, \citenamefont
  {Even}, \citenamefont {Ferrer}, \citenamefont {Giacoppo}, \citenamefont
  {G{\"o}tz} \emph {et~al.}}]{chhetri17}%
  \BibitemOpen
  \bibfield  {author} {\bibinfo {author} {\bibfnamefont {P.}~\bibnamefont
  {Chhetri}}, \bibinfo {author} {\bibfnamefont {D.}~\bibnamefont {Ackermann}},
  \bibinfo {author} {\bibfnamefont {H.}~\bibnamefont {Backe}}, \bibinfo
  {author} {\bibfnamefont {M.}~\bibnamefont {Block}}, \bibinfo {author}
  {\bibfnamefont {B.}~\bibnamefont {Cheal}}, \bibinfo {author} {\bibfnamefont
  {C.~E.}\ \bibnamefont {D{\"u}llmann}}, \bibinfo {author} {\bibfnamefont
  {J.}~\bibnamefont {Even}}, \bibinfo {author} {\bibfnamefont {R.}~\bibnamefont
  {Ferrer}}, \bibinfo {author} {\bibfnamefont {F.}~\bibnamefont {Giacoppo}},
  \bibinfo {author} {\bibfnamefont {S.}~\bibnamefont {G{\"o}tz}},  \emph
  {et~al.},\ }\href {\doibase 10.1140/epjd/e2017-80122-x} {\bibfield  {journal}
  {\bibinfo  {journal} {Europ. Phys. J. D}\ }\textbf {\bibinfo {volume} {71}},\
  \bibinfo {pages} {195} (\bibinfo {year} {2017})}\BibitemShut {NoStop}%
\end{thebibliography}%

\clearpage
\afterpage{\onecolumngrid
\renewcommand{\doublerulesep}{1pt}
\begin{longtable}[e]{@{\extracolsep{\fill}}lllllll}
\caption{Einstein $A$-coefficients for electric-dipole (E1) transitions in lutetium, as calculated by using CI+MBPT wavefunctions and experimental transition energies. We compare our calculated values with experimentally derived $A$-coefficients ($A_{\mathrm{NIST}}$) tabulated in ref \cite{nistasd54}.}\\
\label{tab:Lu_transitions}\\
\hline\hline
\multicolumn{2}{c}{Upper level}	& \multicolumn{2}{c}{Lower Level}	&\multicolumn{1}{c}{Transition energy (cm$^{-1}$)}	&\multicolumn{1}{c}{$A_{\mathrm{CI+MBPT}}$(s$^{-1}$)}	&\multicolumn{1}{c}{$A_{\mathrm{NIST}}$ (s$^{-1}$)}\\
\hline
\endfirsthead
\hline
\multicolumn{2}{c}{Upper level}	& \multicolumn{2}{c}{Lower Level}	&\multicolumn{1}{c}{Transition energy (cm$^{-1}$)}	&\multicolumn{1}{c}{$A_{\mathrm{CI+MBPT}}$(s$^{-1}$)}	&\multicolumn{1}{c}{$A_{\mathrm{NIST}}$ (s$^{-1}$)}\\
\hline
\endhead
\hline
\multicolumn{6}{c}{\textit{Continued next page\ldots}}\\
\endfoot
\hline\hline
\endlastfoot
5d6s6p	&$^4F^o_{3/2}$ 	&6s$^2$5d	&$^2D_{3/2}$		&17427	&2.54$\times 10^6$ 	&1.73$\times 10^6$\\
5d6s6p	&$^4F^o_{3/2}$ 	&6s$^2$5d	&$^2D_{5/2}$		&15433	&9.07$\times 10^4$ 	&7.2$\times 10^4$\\
5d6s6p	&$^4F^o_{5/2}$ 	&6s$^2$5d	&$^2D_{3/2}$		&18505	&1.81$\times 10^6$ 	&1.20$\times 10^6$\\
5d6s6p	&$^4F^o_{5/2}$ 	&6s$^2$5d	&$^2D_{5/2}$		&16511	&1.49$\times 10^6$ 	&9.2$\times 10^5$\\
5d6s6p	&$^4D^o_{1/2}$ 	&6s$^2$5d	&$^2D_{3/2}$		&20762	&1.06$\times 10^6$ 	&9.0$\times 10^5$\\
5d6s6p	&$^4D^o_{3/2}$ 	&6s$^2$5d	&$^2D_{3/2}$		&21195	&5.58$\times 10^4$ 	&1.13$\times 10^5$\\
5d6s6p	&$^4D^o_{3/2}$ 	&6s$^2$5d	&$^2D_{5/2}$		&19201	&2.57$\times 10^5$ 	&2.5$\times 10^5$\\
5d6s6p	&$^2D^o_{5/2}$ 	&6s$^2$5d	&$^2D_{3/2}$		&21462	&4.01$\times 10^6$ 	&3.15$\times 10^6$\\
5d6s6p	&$^2D^o_{5/2}$ 	&6s$^2$5d	&$^2D_{5/2}$		&19468	&1.15$\times 10^7$ 	&9.1$\times 10^6$\\
5d6s6p	&$^2D^o_{3/2}$ 	&6s$^2$5d	&$^2D_{3/2}$		&22125	&3.22$\times 10^7$ 	&2.26$\times 10^7$\\
5d6s6p	&$^2D^o_{3/2}$ 	&6s$^2$5d	&$^2D _{5/2}$	&20131	&7.35$\times 10^4$ 	&3.3$\times 10^4$\\
5d6s6p	&$^4D^o_{5/2}$ 	&6s$^2$5d	&$^2D_{3/2}$		&22222	&3.37$\times 10^5$ 	&1.28$\times 10^5$\\
5d6s6p	&$^4D^o_{5/2}$ 	&6s$^2$5d	&$^2D_{5/2}$		&20228	&2.21$\times 10^6$ 	&9.7$\times 10^5$\\
6s$^2$7s		&$^2S_{1/2}$ 	&6s$^2$6p	&$^2P^o_{1/2}$	&19990	&2.78$\times 10^7$ 	&3.20$\times 10^7$\\
6s$^2$7s		&$^2S_{1/2}$ 	&6s$^2$6p	&$^2P^o_{3/2}$	&16650	&4.86$\times 10^7$ 	&4.9$\times 10^7$\\
5d6s6p	&$4P^o_{3/2}$ 	&6s$^2$5d	&$^2D_{3/2}$ 	&24308	&1.07$\times 10^6$ 	&4.3$\times 10^5$\\
5d6s6p	&$4P^o_{3/2}$ 	&6s$^2$5d	&$^2D_{5/2}$ 	&22314	&9.13$\times 10^4$ 	&8.8$\times 10^4$\\
5d6s6p	&$4P^o_{5/2}$ 	&6s$^2$5d	&$^2D_{3/2}$ 	&25192	&3.98$\times 10^6$ 	&2.21$\times 10^6$\\
5d6s6p	&$4P^o_{5/2}$ 	&6s$^2$5d	&$^2D_{5/2}$ 	&23198	&1.22$\times 10^6$ 	&6.6$\times 10^5$\\
5d6s6p	&$^2F^o_{5/2}$ 	&6s$^2$5d	&$^2D_{3/2}$ 	&28020	&1.05$\times 10^8$ 	&6.9$\times 10^7$\\
5d6s6p	&$^2F^o_{5/2}$ 	&6s$^2$5d	&$^2D_{5/2}$ 	&26026	&3.52$\times 10^7$ 	&2.65$\times 10^7$\\
5d6s6p	&$^2F^o_{7/2}$ 	&6s$^2$5d	&$^2D_{5/2}$ 	&27493	&1.18$\times 10^7$ 	&6.4$\times 10^6$\\
5d6s6p	&$^2F^o_{5/2}$ 	&6s$^2$5d	&$^2D_{3/2}$ 	&30184	&1.31$\times 10^8$ 	&1.85$\times 10^8$\\
6s$^2$6d		&$^2D_{3/2}$ 	&6s$^2$6p	&$^2P^o_{3/2}$ 	&24066	&1.70$\times 10^7$ 	&1.68$\times 10^7$\\
6s$^2$6d		&$^2D_{5/2}$ 	&6s$^2$6p	&$^2P^o_{3/2}$ 	&24237	&9.08$\times 10^7$ 	&8.9$\times 10^7$\\
5d6s6p	&$^2F^o_{7/2}$ 	&6s$^2$5d	&$^2D_{5/2}$ 	&29757	&2.77$\times 10^8$ 	&2.44$\times 10^8$\\
6s$^2$5f		&$^2F^o_{5/2}$ 	&6s$^2$5d	&$^2D_{3/2}$ 	&36633	&1.87$\times 10^7$ 	&2.32$\times 10^7$\\
6s$^2$5f		&$^2F^o_{7/2}$ 	&6s$^2$5d	&$^2D_{5/2}$ 	&34650	&2.52$\times 10^7$ 	&2.31$\times 10^7$\\
\end{longtable}
\twocolumngrid
}

\afterpage{\onecolumngrid
\renewcommand{\doublerulesep}{1pt}
\begin{longtable}[e]{@{\extracolsep{\fill}}lllllll}
\caption{Einstein $A$-coefficients for electric-dipole (E1) transitions in lawrencium, as calculated within the CI+MBPT framework. Branching ratios for each transition are also shown. We estimate the uncertainty in the coefficients as 40\% (see text).}\\
\label{tab:E1_transitions}\\
\hline\hline
\multicolumn{2}{c}{Upper level}	& \multicolumn{2}{c}{Lower Level}	&\multicolumn{1}{c}{Transition energy (cm$^{-1}$)}	&\multicolumn{1}{c}{A(s$^{-1}$)}	&\multicolumn{1}{c}{Branching ratio}	\\
\hline
\endfirsthead
\hline\hline
\multicolumn{2}{c}{Upper level}	& \multicolumn{2}{c}{Lower Level}	&\multicolumn{1}{c}{Transition energy (cm$^{-1}$)}		&\multicolumn{1}{c}{A(s$^{-1}$)}	&\multicolumn{1}{c}{Branching ratio}	\\
\hline
\endhead
\hline
\multicolumn{7}{c}{\textit{Continued next page\ldots}}\\
\endfoot
\hline\hline
\endlastfoot
7s$^2$6d  &$^2D_{3/2}$	   &7s$^2$7p  &$^2P^o_{1/2}$	&712	&1.26$\times 10^{3}$	&1	\\
\\
7s$^2$8s  &$^2S_{1/2}$	   &7s$^2$7p  &$^2P^o_{3/2}$	&11878	&3.57$\times 10^{7}$	&0.52	\\
7s$^2$8s  &$^2S_{1/2}$	   &7s$^2$7p  &$^2P^o_{1/2}$	&20485	&3.31$\times 10^{7}$	&0.48	\\
\\
7s6d$^2$  &$^4F_{3/2}$	   &7s7p6d   &$^4F^o_{5/2}$	&1453	&3.59$\times 10^{3}$	&2.92$\times 10^{-3}$	\\
7s6d$^2$  &$^4F_{3/2}$	   &7s7p6d   &$^4F^o_{3/2}$	&3757	&2.58$\times 10^{5}$	&0.24	\\
7s6d$^2$  &$^4F_{3/2}$	   &7s$^2$7p  &$^2P^o_{3/2}$	&16136	&6.73$\times 10^{3}$	&0.01	\\
7s6d$^2$  &$^4F_{3/2}$	   &7s$^2$7p  &$^2P^o_{1/2}$	&24742	&7.96$\times 10^{5}$	&0.75	\\
\\
7s7p$^2$  &$^4P_{1/2}$	   &7s7p6d   &$^4F^o_{3/2}$	&4396	&2.27$\times 10^{3}$	&8.06$\times 10^{-5}$	\\
7s7p$^2$  &$^4P_{1/2}$	   &7s$^2$7p  &$^2P^o_{3/2}$	&16775	&3.22$\times 10^{6}$	&0.11	\\
7s7p$^2$  &$^4P_{1/2}$	   &7s$^2$7p  &$^2P^o_{1/2}$	&25381	&2.51$\times 10^{7}$	&0.89	\\
\\
7s6d$^2$  &$^4F_{5/2}$	   &7s7p6d   &$^4F^o_{5/2}$	&2876	&9.60$\times 10^{4}$	&0.85	\\
7s6d$^2$  &$^4F_{5/2}$	   &7s7p6d   &$^4F^o_{3/2}$	&5180	&1.36$\times 10^{4}$	&0.12	\\
7s6d$^2$  &$^4F_{5/2}$	   &7s$^2$7p  &$^2P^o_{3/2}$	&17559	&3.00$\times 10^{3}$	&0.03	\\
\\
7s$^2$7d  &$^2D_{3/2}$	   &7s$^2$8p  &$^2P^o_{3/2}$	&273	&2.51$\times 10^{2}$	&3.38$\times 10^{-6}$	\\
7s$^2$7d  &$^2D_{3/2}$	   &7s$^2$8p  &$^2P^o_{1/2}$	&1585	&1.73$\times 10^{5}$	&1.39$\times 10^{-3}$	\\
7s$^2$7d  &$^2D_{3/2}$	   &7s7p6d   &$^4D^o_{3/2}$	&1772	&2.43$\times 10^{4}$	&3.26$\times 10^{-4}$	\\
7s$^2$7d  &$^2D_{3/2}$	   &7s7p6d   &Odd ($J = 1/2$)	&2693	&7.09$\times 10^{5}$	&0.01	\\
7s$^2$7d  &$^2D_{3/2}$	   &7s7p6d   &$^4F^o_{5/2}$	&5291	&3.30$\times 10^{3}$	&4.44$\times 10^{-5}$	\\
7s$^2$7d  &$^2D_{3/2}$	   &7s7p6d   &$^4F^o_{3/2}$	&7585	&2.52$\times 10^{2}$	&3.40$\times 10^{-6}$	\\
7s$^2$7d  &$^2D_{3/2}$	   &7s$^2$7p  &$^2P^o_{3/2}$	&19974	&1.21$\times 10^{7}$	&0.16	\\
7s$^2$7d  &$^2D_{3/2}$	   &7s$^2$7p  &$^2P^o_{1/2}$	&28580	&6.14$\times 10^{7}$	&0.83	\\
\\
7s$^2$7d  &$^2D_{5/2}$     &7s7p6d    &$^4D^o_{5/2}$ &18  	&1.72$\times 10^{-3}$	&3.18$\times 10^{-11}$\\
7s$^2$7d  &$^2D_{5/2}$	   &7s7p6d   &$^4F^o_{7/2}$	&151	&5.44$\times 10^{-1}$	&1.01$\times 10^{-8}$\\
7s$^2$7d  &$^2D_{5/2}$	   &7s$^2$8p  &$^2P^o_{3/2}$	&419	&5.47$\times 10^{3}$	&1.01$\times 10^{-4}$	\\
7s$^2$7d  &$^2D_{5/2}$	   &7s7p6d   &$^4D^o_{3/2}$	&1917	&1.97$\times 10^{5}$	&3.66$\times 10^{-3}$	\\
7s$^2$7d  &$^2D_{5/2}$	   &7s7p6d   &$^4F^o_{5/2}$	&5436	&3.97$\times 10^{3}$	&7.32$\times 10^{-5}$	\\
7s$^2$7d  &$^2D_{5/2}$	   &7s7p6d   &$^4F^o_{3/2}$	&7740	&4.18$\times 10^{3}$	&7.73$\times 10^{-5}$	\\
7s$^2$7d  &$^2D_{5/2}$	   &7s$^2$7p  &$^2P^o_{3/2}$	&20119	&5.39$\times 10^{7}$	&0.99	\\
\\
7s$^2$7p  &$^2P^o_{3/2}$	   &7s$^2$6d  &$^2D_{5/2}$	&3355	&3.13$\times 10^{5}$	&0.49	\\
7s$^2$7p  &$^2P^o_{3/2}$	   &7s$^2$6d  &$^2D_{3/2}$	&7894	&3.01$\times 10^{5}$	&0.51	\\
\\
7s7p6d    &$^4F^o_{3/2}$	   &7s$^2$8s  &$^2S_{1/2}$	&501	&1.53$\times 10^{-1}$	&5.74$\times 10^{-8}$	\\
7s7p6d    &$^4F^o_{3/2}$	   &7s$^2$6d  &$^2D_{5/2}$	&15734	&1.56$\times 10^{5}$	&0.02	\\
7s7p6d    &$^4F^o_{3/2}$	   &7s$^2$6d  &$^2D_{3/2}$	&20273	&8.80$\times 10^{6}$	&0.98	\\
\\
7s7p6d    &$^4F^o_{5/2}$	   &7s$^2$6d  &$^2D_{5/2}$	&18038	&3.26$\times 10^{6}$	&0.19	\\
7s7p6d    &$^4F^o_{5/2}$	   &7s$^2$6d  &$^2D_{3/2}$	&22577	&1.35$\times 10^{7}$	&0.81	\\
\\
7s7p6d    &Odd ($J = 1/2$)	   &7s7p$^2$  &$^4P_{1/2}$	&506	&5.39$\times 10^{2}$	&2.11$\times 10^{-5}$	\\
7s7p6d    &Odd ($J = 1/2$)	   &7s6d$^2$  &$^4F_{3/2}$	&987	&1.20$\times 10^{4}$	&4.70$\times 10^{-4}$	\\
7s7p6d    &Odd ($J = 1/2$)	   &7s$^2$8s  &$^2S_{1/2}$	&5402	&3.58$\times 10^{6}$	&0.14	\\
7s7p6d    &Odd ($J = 1/2$)	   &7s$^2$6d  &$^2D_{3/2}$	&25175	&2.21$\times 10^{7}$	&0.86	\\
\\
7s7p6d    &$^4D^o_{3/2}$	   &7s6d$^2$  &$^4F_{5/2}$	&643	&2.41$\times 10^{3}$	&1.19$\times 10^{-4}$\\
7s7p6d    &$^4D^o_{3/2}$	   &7s7p$^2$  &$^4P_{1/2}$	&1427	&7.41$\times 10^{3}$	&3.66$\times 10^{-4}$\\
7s7p6d    &$^4D^o_{3/2}$	   &7s6d$^2$  &$^4F_{3/2}$	&2066	&9.36$\times 10^{3}$	&4.62$\times 10^{-4}$\\
7s7p6d    &$^4D^o_{3/2}$	   &7s$^2$8s  &$^2S_{1/2}$	&6324	&3.22$\times 10^{6}$	&0.16	\\
7s7p6d    &$^4D^o_{3/2}$	   &7s$^2$6d  &$^2D_{5/2}$	&21557	&2.59$\times 10^{6}$	&0.13	\\
7s7p6d    &$^4D^o_{3/2}$	   &7s$^2$6d  &$^2D_{3/2}$	&26283	&1.44$\times 10^{7}$	&0.71	\\
\\
7s$^2$8p  &$^2P^o_{1/2}$	   &7s7p$^2$  &$^4P_{1/2}$	&1615	&5.95$\times 10^{2}$	&8.43$\times 10^{-5}$	\\
7s$^2$8p  &$^2P^o_{1/2}$	   &7s6d$^2$  &$^4F_{3/2}$	&2253	&5.27$\times 10^{4}$	&0.007	\\
7s$^2$8p  &$^2P^o_{1/2}$	   &7s$^2$8s  &$^2S_{1/2}$	&6511	&6.42$\times 10^{6}$	&0.91	\\
7s$^2$8p  &$^2P^o_{1/2}$	   &7s$^2$6d  &$^2D_{3/2}$	&26283	&5.95$\times 10^{5}$	&0.08	\\
\\
7s$^2$8p  &$^2P^o_{3/2}$	   &7s6d$^2$  &$^4F_{5/2}$	&2141	&8.12$\times 10^{3}$	&5.30$\times 10^{-4}$\\
7s$^2$8p  &$^2P^o_{3/2}$	   &7s7p$^2$  &$^4P_{1/2}$	&2926	&2.73$\times 10^{4}$	&1.79$\times 10^{-3}$\\
7s$^2$8p  &$^2P^o_{3/2}$	   &7s6d$^2$  &$^4F_{3/2}$	&3564	&1.35$\times 10^{4}$	&8.82$\times 10^{-4}$\\
7s$^2$8p  &$^2P^o_{3/2}$	   &7s$^2$8s  &$^2S_{1/2}$	&7822	&1.14$\times 10^{7}$	&0.75	\\
7s$^2$8p  &$^2P^o_{3/2}$	   &7s$^2$6d  &$^2D_{5/2}$	&23055	&6.14$\times 10^{4}$	&3.80$\times 10^{-3}$\\
7s$^2$8p  &$^2P^o_{3/2}$	   &7s$^2$6d  &$^2D_{3/2}$	&27595	&3.76$\times 10^{6}$	&0.24	\\
\\
7s7p6d    &$^4F^o_{7/2}$	   &7s6d$^2$  &$^4F_{7/2}$	&284	&1.135$\times 10^2$	                &2.39$\times 10^{-5}$\\
7s7p6d    &$^4F^o_{7/2}$	   &7s6d$^2$  &$^4F_{5/2}$	&2409	&1.17$\times 10^{4}$	&2.00$\times 10^{-3}$\\
7s7p6d    &$^4F^o_{7/2}$	   &7s$^2$6d  &$^2D_{5/2}$	&23323	&4.74$\times 10^{6}$	&0.998	\\
\\
7s7p6d    &$^4D^o_{5/2}$	   &7s6d$^2$  &$^4F_{7/2}$	&417	&1.07$\times 10^{2}$	&3.48$\times 10^{-6}$\\
7s7p6d    &$^4D^o_{5/2}$	   &7s$^2$7d  &$^2D_{3/2}$	&127	&5.53$\times 10^{-3}$	&5.79$\times 10^{-11}$\\
7s7p6d    &$^4D^o_{5/2}$	   &7s6d$^2$  &$^4F_{5/2}$	&2542	&1.14$\times 10^{4}$	&3.69$\times 10^{-4}$\\
7s7p6d    &$^4D^o_{5/2}$	   &7s6d$^2$  &$^4F_{3/2}$	&3965	&4.99$\times 10^{3}$	&1.62$\times 10^{-4}$\\
7s7p6d    &$^4D^o_{5/2}$	   &7s$^2$6d  &$^2D_{5/2}$	&23456	&2.03$\times 10^{7}$	&0.66\\
7s7p6d    &$^4D^o_{5/2}$	   &7s$^2$6d  &$^2D_{3/2}$	&27966	&1.05$\times 10^{7}$	&0.34\\
\end{longtable}
\twocolumngrid
}
\end{document}